\documentclass[smallextended]{svjour3}   
\pdfoutput=1 

\usepackage{citesort}
\usepackage{graphicx}
\usepackage{bm}
\usepackage{amsmath}
\usepackage{amssymb}

\newenvironment{remarks}[1][\textit{Remarks}:]{\begin{trivlist}
\item[\hskip \labelsep { #1}]}{\end{trivlist}}
\newenvironment{remarkk}[1][\textit{Remark}:]{\begin{trivlist}
\item[\hskip \labelsep { #1}]}{\end{trivlist}}

\newcommand{\sref}[1]{Section \ref{#1}}
\newcommand{\fref}[1]{Figure \ref{#1}}
\newcommand{\aref}[1]{Appendix \ref{#1}}

\newcommand{\rrangle}{\right \rangle}
\newcommand{\llangle}{\left \langle}
\newcommand{\rt}{\textrm{right}}
\newcommand{\lt}{\textrm{left}}

\journalname{J. Stat. Phys.}

\begin{document}

\title{Large deviations conditioned on large deviations I: Markov chain and Langevin equation.}

\author{Bernard Derrida   \and
        Tridib Sadhu
}

\institute{Bernard Derrida \at
              Coll\`{e}ge de France, 11 place Marcelin Berthelot, 75231 Paris Cedex 05 - France.\\
              bernard.derrida@college-de-france.fr
           \and
          Tridib Sadhu  \at
              Coll\`{e}ge de France, 11 place Marcelin Berthelot, 75231 Paris Cedex 05 - France.\\
          Tata Institute of Fundamental Research, Homi Bhabha Road, Mumbai 400005.\\
          tridib@theory.tifr.res.in
}

\maketitle

\begin{abstract}
We present a systematic analysis of stochastic processes conditioned on an empirical measure $Q_T$ defined in a time interval $[0,T]$ for large $T$. We build our analysis starting from a discrete time Markov chain. Results for a continuous time Markov process and Langevin dynamics are derived as limiting cases. We show how conditioning on a value of $Q_T$ modifies the dynamics. For a Langevin dynamics with weak noise, we introduce conditioned large deviations functions and calculate them using either a WKB method or a variational formulation.
This allows us, in particular, to calculate the typical trajectory and the fluctuations around this optimal trajectory when conditioned on a certain value of $Q_T$.

\end{abstract}

\section{Introduction}
Understanding the frequency of rare  events and the dynamical trajectories
which generate them has become an important  field of research in many
physical situations including protein folding \cite{Mey2014}, chemical reactions \cite{Delarue2017,Dykman1994},
atmospheric activities \cite{Laurie2015}, glassy systems \cite{Garrahan2007,Garrahan2009}, disordered media \cite{Dorlas2001}, \textit{etc.}.
From the mathematical point of view, the statistical properties of rare
events are characterized by large deviations functions \cite{dembo2009,Varadhan1966,Varadhan2003,Donsker19752,Hollander2008,TOUCHETTE2009,ellis1985,Derrida2007,Hurtado2014}.
In physics, a particular interest for large deviations functions arose
in the context of non-equilibrium statistical physics from the
discovery of the fluctuation theorem \cite{Kurchan1998,Gallavotti1995,Lebowitz1999} where the rare event consists
in observing an atypical value of a current over a long time window.  They also had been used for a long
time to study stochastic dynamical systems in a weak noise limit \cite{Freidlin2012,Graham1985,Graham1973} or
extended systems when the system size becomes large \cite{Bertini2014,Derrida2007,Bertini2001}.

One of the simplest questions one may ask about the large deviations
functions is to consider an empirical measure $Q_T$ of the form
\begin{equation}
Q_T= \int_0^T  dt\,f({ C}_t)
\end{equation}
where $f({ C}_t)$ is a function of the configuration ${ C}_t$ of a stochastic (or a chaotic) system at
time $t$ and to try to determine the probability that this empirical
measure takes any atypical value $q\,T$. For large $T$, the large
deviations function $\phi(q)$ is then simply defined by \cite{Varadhan1984,Donsker19752,Maes2008,Maes20082,Bodineau2004,Derrida2007,Bertini2005Current,Hurtado2014,Hurtado2010}
\begin{equation}
\textrm{ Prob}(Q_T=q T) \sim e^{-T \, \phi(q)}  \qquad   \textrm{for large } T
\end{equation}
(Here the precise meaning of the symbol $\sim$ is that $\lim_{T\rightarrow \infty}\frac{1}{T}\log \textrm{Prob}(q T)=-\phi(q)$, and this will be used throughout this article.) A rather common situation is when $\phi(q)$ vanishes at a single value
$q^*$ of $q$ (the most likely value of $q$) and where $\phi(q) >0$ for $q
\neq q^*$. The main question we try to address in the present paper is what are the
dominant trajectories of a stochastic process which  contribute to this large deviations function and
how to describe their effective dynamics.
In particular, we want to understand how to predict the probability 
$P_t({ C}|Q_T=q\, T)$ of finding the system in a configuration ${C}$ at an arbitrary time $t$, conditioned on a certain value of $Q_T$.

A very related approach \cite{Jack2010,Jack2015,Chetrite2013,Chetrite2015,TOUCHETTE2017,Maes2008,Maes20082,Lecomte20072,Hartmann2012,Bertini2015} (what we will call the canonical approach)
consists in weighting  all the events by an exponential of $Q_T$ and  to
try to determine the probability
\begin{equation}
P_t^{(\lambda)} (C)=\frac{\int dQ \, e^{\lambda Q} P_t(C,Q)}{\sum_{C'}\int dQ \, e^{\lambda Q}P_t(C',Q)}
\end{equation}
where $P_t(C,Q)$ is the joint probability of configuration $C$ at time $t$ and the observable $Q_T$ to take value $Q$ given the system in its steady state. This is in contrast to the previous case (where $Q_T$ was fixed and that
we call the microcanincal case). As we shall see (in particular, in \sref{sec:Markov}
 and \aref{app:equivalence})  these canonical and microcanonical ensembles are
related in the usual way in the large $T$ limit (which plays here the same
role as the thermodynamic limit in standard statistical mechanics).

\begin{figure}
\centering \includegraphics[width=0.9\textwidth]{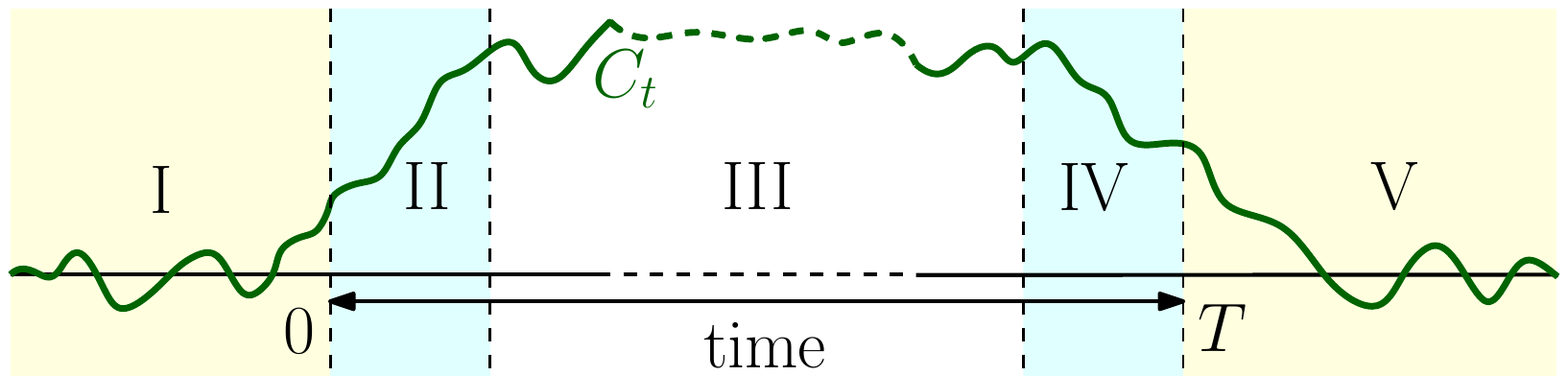}
\caption{A schematic of a time evolution of a Markov process $C_t$ when conditioned on an empirical observable $Q_T$ measured in a large time interval $[0,T]$. Different regions denote different parts of the evolution: (I) $t<0$, (II) $t\ge 0$ but small, (III) $t$ and $T-t$ both large, (IV) $T-t> 0$ but small, and (V) $t\ge T$. \label{fig:fig1}}
\end{figure}

 Our paper will start by reviewing and extending some known aspects of
the large deviations function for Markov processes and for the Langevin
equation  (see \sref{sec:Markov} and \sref{sec:langevin}). In the large $T$ limit,
one has to distinguish five  regions (see \fref{fig:fig1}) for which we calculate
how the measure and the dynamics are modified by the conditioning on
$Q_T$. Then, we will consider the Langevin equation in the weak noise limit, first
using a Wentzel-Kramers-Brillouin (WKB) approach \cite{Landau} (\sref{sec:ldf variational big}) and a variational approach (\sref{sec:variational})
based on the search of an optimal path which minimizes an action. This
will allow in particular to obtain the equation followed by
the optimal trajectory under conditioning. Lastly we will see in \sref{sec:effective noise} that the effect of conditioning is to
break causality in the sense that a trajectory becomes correlated to the noise in the future.

\section{Markov Process \label{sec:Markov}}

For large $T$, a schematic time evolution of a Markovian stochastic system conditioned to take a certain value of $Q_T$ is shown in \fref{fig:fig1} where one has to consider five regions. The system starts from a typical configuration far in the past, and evolves to a quasi-stationary regime (region III in \fref{fig:fig1}), and finally relaxes to the typical state of the unconditioned dynamics. One knows \cite{Jack2010,Jack2015,Chetrite2013,Chetrite2015,TOUCHETTE2017,Lecomte20072,Garrahan2007,Garrahan2009,Evans2004} how to describe the effective dynamics in the quasi-stationary regime. For a Markov chain, the effective dynamics in region III is known to remain Markovian with transition rate which can be expressed in terms of the largest eigenvalue and eigenvectors of the tilted Markov matrix. This connection between conditioned dynamics and a biased ensemble appeared earlier in many contexts: Doob's h-transformation \cite{Stroock}, Donsker-Varadhan theory of large deviations \cite{Donsker19752}, rare events problems \cite{Jack2015,Chetrite2013,Chetrite2015,TOUCHETTE2017,Maes2008,Maes20082,Hartmann2012,Lecomte20072,Hirschberg2015,Schutz2016,Popkov2011,Popkov2010,Evans2004}, kinetically constrained models \cite{Garrahan2007,Garrahan2009}, optimal control theory \cite{Fleming1992,Chetrite20152}, and even in Quantum systems \cite{Carollo2018}. In this section, we give a simple derivation of the effective dynamics which extends to the five regions of \fref{fig:fig1}, the earlier results known in the quasi-stationary regime.

\subsection{The tilted matrix \label{sec:deform}}

We focus here our discussion on a discrete time irreducible Markov process on a finite set of configurations. This Markov process is specified by the probability $M_0(C',C)$ that the system jumps from configuration $C$ to $C'$ in one time step. As we will see later, the continuous time Markov process and the Langevin dynamics can be obtained as limiting cases.

For this discrete time Markov process, we want to condition on a general empirical measure
\begin{equation}
Q_T= \sum_{t=0}^{T-1} f({\cal C}_t) + \sum_{t=0}^{T-1}g({\cal C}_{t+1},{\cal C}_{t})
\label{eq:QT}
\end{equation}
where $f$ and $g$ are arbitrary functions of the configurations. 
For example, by choosing $f(C) = \delta_{C,C_a} $ and $g(C',C)=0$, the observable  $Q_T$ represents the total time spent in a particular configuration $C_a$. Another choice $f(C)=0$ and $g(C',C)=\delta_{C',C_b}\delta_{C,C_a}$ would count the total number of jumps from configuration $C_a$ to configuration $C_b$.

Our goal is to describe the dynamics conditioned on a certain value of $Q_T$ for large $T$. In particular, we want to know what is the conditional probability $\mathcal{P}_t(C\vert Q_T)$ for the system to be in a configuration $C$ at an arbitrary time $t$ when conditioned on the observable $Q_T$ defined by \eqref{eq:QT}.

Let us first analyze the special case $t=T$. If we define the joint probability $P_T(C,Q\vert C_0)$ for the system to be in a configuration $C$ at time $T$ and that the observable $Q_T$ defined by \eqref{eq:QT} takes value $Q$ given its initial configuration $C_0$ at time $0$, it satisfies a recursion relation:
\begin{equation}
P_T(C,Q \vert C_0)=\sum_{C'}M_0(C,C')P_{T-1}(C',Q-f(C')-g(C,C') \vert C_0)
\label{eq:joint P}
\end{equation}
Then, it is easy to see that the generating function defined by
\begin{equation}
G_{T}^{(\lambda)}(C \vert C_0)=\int dQ \;e^{\lambda Q}P_{T}(C,Q \vert C_0)
\label{eq:G}
\end{equation}
satisfies 
\begin{equation}
G_{T}^{(\lambda)}(C \vert C_0)=\sum_{C'}M_{\lambda}(C,C')G_{T-1}^{(\lambda)}(C'\vert C_0)
\label{eq:GT recursion}
\end{equation}
where
\begin{equation}
M_\lambda(C,C')= M_0(C,C') e^{\lambda\left[ f(C') + g(C,C')\right]}
\label{eq:tilted discrete}
\end{equation}
is the tilted matrix \cite{Lebowitz1999,Derrida2007,Chetrite2013,Lecomte20072,Garrahan2009,Roche2004,Hirschberg2015,Popkov2010,Bertini2015}. Therefore, $G_{T}^{(\lambda)}(C \vert C_0)=M_{\lambda}^T(C,C_0)$ is the $(C,C_0)$th element of the matrix $(M_{\lambda})^T$. For large $T$, the matrix elements of $(M_{\lambda})^T$ are dominated by the largest eigenvalue $e^{\mu(\lambda)}$ of $M_{\lambda}$, resulting in
\begin{equation}
G_{T}^{(\lambda)}(C \vert C_0)\simeq e^{T \mu(\lambda)}R_{\lambda}(C)L_{\lambda}(C_0)
\label{eq:main G}
\end{equation}
where $R_{\lambda}(C)$ and $L_{\lambda}(C)$ are the associated right and left eigenvectors, respectively. For the prefactor in \eqref{eq:main G} to be correct the eigenvectors must be normalized with $\sum_C R_{\lambda}(C)L_{\lambda}(C)=1$. 

\begin{remarks}~\\
\begin{enumerate}
\item
It follows from (\ref{eq:G},\,\ref{eq:main G}) that the cumulants of $Q_T$, for large $T$, can be obtained from the derivatives of $\mu(\lambda)$ at $\lambda=0$, and that $\lim_{T\rightarrow \infty}\frac{1}{T}\log\langle e^{\lambda Q_T}\rangle=\mu(\lambda)$.
\item
The Perron-Frobenius theorem \cite{VANKAMPEN2007193} ensures that the largest eigenvalue of $M_{\lambda}$ is positive and non-degenerate, and all components of the associated right and left eigenvectors are positive. For non-zero $\lambda$, the tilted matrix $M_\lambda$ is, in general, not Markovian (because $\sum_{C'}M_\lambda(C',C)\ne 1$) and non-Hermitian.
\item
In the case $\lambda=0$, the largest eigenvalue is $1$, with $L_0(C)=1$, and $R_0(C)$ is the steady state probability distribution of the Markov process $M_0$.
\end{enumerate}
\end{remarks}
 
\subsection{Ensemble equivalence} 
From \eqref{eq:G} and \eqref{eq:main G}, one can see by a saddle point calculation that for large $T$
\begin{equation}
P_T(C,Q=qT\vert C_0)\simeq e^{-T \phi(q)}\sqrt{\frac{\phi''(q)}{2\pi T}}R_{\phi'(q)}(C)L_{\phi'(q)}(C_0)
\label{eq:PT large dev}
\end{equation}
where the large deviation function $\phi(q)$ and the eigenvalue $e^{\mu(\lambda)}$ of the matrix $M_{\lambda}$ are related by a Legendre transformation
\begin{equation}
\mu(\lambda)=\lambda q-\phi(q)\qquad \mbox{with } \qquad\lambda=\phi'(q)
\label{eq:legendre}
\end{equation}
We see from \eqref{eq:PT large dev} that, for large $T$, the conditional distribution of $C$ at the final time is given by
\begin{equation}
\mathcal{P}_{T}(C\vert Q=qT)= \frac{P_T(C,Q=qT\vert C_0)}{\sum_{C'}P_T(C',Q=qT\vert C_0)}\simeq \frac{R_{\phi'(q)}(C)}{\sum_{C'} R_{\phi'(q)}(C')}
\label{eq:result P cond}
\end{equation}
This shows that the initial condition $C_0$ is forgotten at large $T$. Therefore, we leave out the reference to $C_0$ in our notation for the conditional probability. On the other hand, in the $\lambda$-ensemble, using \eqref{eq:main G} one has the probability at the final time
\begin{equation}
P_{T}^{(\lambda)}(C)=\frac{G_T^{(\lambda)}(C\vert C_0)}{\sum_{C'}G_T^{(\lambda)}(C'\vert C_0)}\simeq \frac{R_{\lambda}(C)}{\sum_{C'} R_{\lambda}(C')}
\label{eq:p lambda T}
\end{equation}

Comparing \eqref{eq:result P cond} and \eqref{eq:p lambda T} we see that the conditional probability $\mathcal{P}_{t=T}(C\vert Q=qT)$ for large $T$ can be obtained from the probability $P_{T}^{(\lambda)}(C)$ by substituting $\lambda=\phi'(q)$.  This shows that, for large $T$, the two ensembles are equivalent: fixing the value of $Q_T$ or weighting the events by a factor $e^{\lambda Q_T}$ lead to the same distribution of the final configuration $C$.  The former is an analogue of the micro-canonical ensemble with fixed $Q_T$ and the latter is its canonical counterpart defined by the conjugate variable $\lambda$.

\begin{remarkk}
For an irreducible Markov process on a finite configuration space, the spectral gap between the largest and the second largest eigenvalues   is non-zero. Moreover, the functions $\phi(q)$ and $\mu(\lambda)$ are analytic and convex, and the equivalence \eqref{eq:legendre} is assured. This may not be the case for systems with infinite configurations, where the gap may disappear and large deviations functions could become non-analytic \cite{Bodineau2005,Lecomte20072,Garrahan2007,Garrahan2009,Jack2010,Baek2017,Espigares2013}.
\end{remarkk}

\subsection{The measure conditioned on $Q_T$ \label{sec:cond prob}}
As shown in \aref{app:equivalence}, the equivalence of ensembles holds not only at time $t=T$, but at any time $t$, as long as $T$ is large \cite{Jack2010,Jack2015,Chetrite2013,Chetrite2015,TOUCHETTE2017}. This states that, by generalizing \eqref{eq:p lambda T}, if we define the canonical probability 
\begin{equation}
P_t^{(\lambda)} (C)=\frac{\int dQ \, e^{\lambda Q} P_t(C,Q)}{\sum_{C'}\int dQ \, e^{\lambda Q}P_t(C',Q)}
\label{eq:cond prob 1}
\end{equation}
for any time $t$, then for large $T$,
\begin{equation}
\mathcal{P}_t(C \vert Q_T=qT)\simeq P_t^{(\lambda)}(C)\qquad \mbox{with}\qquad \lambda=\phi'(q)
\label{eq:equivalence 2}
\end{equation}
where $P_t(C,Q)$ is the joint probability of configuration $C$ at time $t$ and the observable $Q_T$ to take value $Q$ given the system in its steady state; $\mathcal{P}_t(C\vert Q)$ is the corresponding conditional probability.

This conditioned measure \eqref{eq:cond prob 1} for large $T$ takes different expressions in the five regions indicated in \fref{fig:fig1}. (A derivation is presented in \aref{app:equivalence} for region II and can be easily extended for other regions.)
\begin{subequations}
\begin{itemize}
\item Region I. $t<0$
\begin{equation}
P_t^{(\lambda)}(C)=\frac{\sum_{C'}L_{\lambda}(C')M_0^{-t}(C',C)R_{0}(C)}{\sum_{C'}L_{\lambda}(C')R_{0}(C')}
\label{eq:prob I}
\end{equation}
\item Region II. $0\le t\ll T$
\begin{equation}
P_t^{(\lambda)}(C)=\frac{\sum_{C'}L_{\lambda}(C)M_{\lambda}^{t}(C,C')R_{0}(C')}{e^{t\mu(\lambda)}\sum_{C'}L_{\lambda}(C')R_{0}(C')}
\label{eq:prob II}
\end{equation}
\item Region III. $1\ll t$ and $T-t\gg 1$
\begin{equation}
P_t^{(\lambda)}(C)=R_{\lambda}(C)L_{\lambda}(C)
\label{eq:prob III}
\end{equation}
\item Region IV. $1\ll t< T$, \textit{i.e.} $T-t=\mathcal{O}(1)$
\begin{equation}
P_t^{(\lambda)}(C)=\frac{\sum_{C'}M_{\lambda}^{T-t}(C',C)R_{\lambda}(C)}{e^{(T-t)\mu(\lambda)}\sum_{C'}R_{\lambda}(C')}
\label{eq:prob IV}
\end{equation}
\item Region V. $T\le t$
\begin{equation}
P_t^{(\lambda)}(C)=\frac{\sum_{C'}M_{0}^{t-T}(C,C')R_{\lambda}(C')}{\sum_{C'}R_{\lambda}(C')}
\label{eq:prob V}
\end{equation}
\end{itemize}
\end{subequations}
To be consistent with the notation of \sref{sec:deform} we denote by $R_0(C)$ the steady state measure of the Markov process $M_0$. Therefore \eqref{eq:p lambda T} is a special case of \eqref{eq:prob IV}. Another special case
\begin{equation}
P_{t=0}^{(\lambda)}(C)=\frac{L_{\lambda}(C)R_0(C)}{\sum_{C'}L_{\lambda}(C')R_0(C')}
\label{eq:P cond t 0}
\end{equation}

\subsection{Time evolution of the conditioned process \label{sec:cond rate}}
Again by a straightforward generalization of the reasoning (see \aref{app:equivalence}), one can show that the equivalence of ensembles holds for the conditioned dynamics as well. In fact, the conditioned dynamics is itself a Markov process \cite{Jack2010,Jack2015,Chetrite2013,Chetrite2015,TOUCHETTE2017}.
For this process, the probability of jump $W_t^{(\lambda)}(C',C)$ from configuration $C$ at $t$ to $C'$ at $t+1$ in the canonical ensemble (events weighted by $e^{\lambda Q_T}$) depends, in general, on time $t$. 
For example, for $t<0$,
\begin{equation*}
W_t^{(\lambda)}(C',C)= {\sum_{C''',C''} M_{\lambda}^{T}(C''',C'') \ M_0^{-t-1}(C'',C') \  M_0(C',C)R_0(C) \over 
 \sum_{C''',C''} M_{\lambda}^{T}(C''',C'') \ M_0^{-t}(C'',C)R_0(C)  }
\end{equation*}
while for $0\le t < T$,
\begin{align*}
W_t^{(\lambda)}(C',C)=\frac{\sum_{C'',C_0}M_{\lambda}^{T-t-1}(C'',C')M_{\lambda}(C',C)M_{\lambda}^t(C,C_0)R_0(C_0)}{\sum_{C'',C_0}M_{\lambda}^{T-t}(C'',C)M_{\lambda}^t(C,C_0)R_0(C_0)}
\end{align*}

For large $T$, the dominant contribution comes from the largest eigenvalue of $M_{\lambda}$, and one gets in the five regions of \fref{fig:fig1}:
\begin{subequations}
\begin{itemize}
\item Region I.
\begin{equation}
W_t^{(\lambda)}(C',C)= {\sum_{C''} L_\lambda(C'') \ M_0^{-t-1}(C'',C') \  M_0(C',C) \over 
 \sum_{C''} L_\lambda(C'') \ M_0^{-t}(C'',C)  }
 \label{eq:tr prob 1}
\end{equation}
\item Region II and III.
\begin{equation}
W_t^{(\lambda)}(C',C)= { L_\lambda(C') \ M_\lambda(C',C)   \over 
e^{\mu(\lambda)}  L_\lambda(C)   } 
\label{eq:tr prob II}
\end{equation}
\item Region IV.
\begin{equation}
W_t^{(\lambda)}(C',C)= {\sum_{C''}  \ M_\lambda^{T-t-1}(C'',C') \  M_\lambda(C',C) \over 
 \sum_{C''}  \ M_\lambda^{T-t}(C'',C) } 
\end{equation}
\item Region V.
\begin{equation}
W_t^{(\lambda)}(C',C)=M_0(C',C)
\label{eq:tr prob 2}
\end{equation}
\end{itemize}
\end{subequations}

Using these expressions for $W_t^{(\lambda)}$ and their corresponding conditioned probability in (\ref{eq:prob I}-\ref{eq:prob IV}), one can check that 
\begin{equation}
P_{t+1}^{(\lambda)}(C')=\sum_{C}W_t^{(\lambda)}(C',C)P_t^{(\lambda)}(C)
\label{eq:Markov W}
\end{equation}

\begin{remarks}
~\\
\begin{enumerate}
\item
We have seen that by deforming the matrix $M_0$
one can condition on two kinds of observables: $f(C_t)$ and $g(C_{t+1},C_t)$ (see \eqref{eq:QT}).
It is not possible to condition on other time correlations, like, 
$Q_T=  \sum_{t=1}^T g( {\cal C}_{t+\tau},{\cal C}_t) $ with $\tau >1$ by simply deforming the matrix $M_0$. One could still define a tilted Markov process  but this would be on a much larger set of configurations since one would need to keep information about $\tau$ consecutive configurations.
\item
In a similar analysis one can describe the time reversed process conditioned on $Q_T$. We define $\mathbb{W}_t^{(\lambda)}(C,C')$ as the transition probability to jump from $C'$ at $t+1$ to $C$ at $t$ in the time reversed process. In all five regions of time, they could be expressed in terms of the corresponding $W_t^{(\lambda)}$ and $P_{t}^{(\lambda)}$ of the forward process.
\begin{equation}
\mathbb{W}_t^{(\lambda)}(C,C')=W_t^{(\lambda)}(C',C)\frac{P_t^{(\lambda)}(C)}{P_{t+1}^{(\lambda)}(C')}
\end{equation}
For example, in the quasi-stationary regime ($1\ll t$ and $T-t\gg 1$),
\begin{equation}
\mathbb{W}_t^{(\lambda)}(C,C')={ M_\lambda(C',C) R_\lambda(C)   \over 
e^{\mu(\lambda)}  R_\lambda(C')   }.
\label{eq:rates time reversed}
\end{equation}

The time reversed process is useful in describing how a fluctuation is created. For example, the fluctuation leading to an atypical configuration can be described by relaxation from the same configuration in the time reversed process \cite{Sadhu2018}.
\end{enumerate}
\end{remarks}

\subsection{A generalization \label{sec:general derivation}}
The above expressions (\ref{eq:prob I}-\ref{eq:prob V}) and (\ref{eq:tr prob 1}-\ref{eq:tr prob 2}) can be extended for a more general observable of the form
\begin{equation}
Q=\sum_t f_t(C_t)+\sum_t g_t(C_{t+1},C_t)
\label{eq:Q general}
\end{equation}
where $f_t(C)$ and $g_t(C',C)$ are arbitrary functions of configurations in a discrete time irreducible Markov process $M_0(C',C)$ on a finite configuration space. To make a clear distinction between the two terms in \eqref{eq:Q general} we shall use $g_t(C,C)=0$. The observable \eqref{eq:QT} is just a particular case of \eqref{eq:Q general} with $f_t(C)=f(C)$ and $g_t(C',C)=g(C',C)$ for $t\in [0,T]$ with large $T$, and both being zero outside this time window. 

We consider that the system started at $t\rightarrow -\infty$ and evolves till $t\rightarrow \infty$, but this can be changed without affecting much of our analysis. One can even generalize to the case when the Markov process $M_0(C',C)$ depends on time.

Using a reasoning similar to that in \aref{app:equivalence}, one can show that in the canonical ensemble where the dynamics is weighted by $e^{\lambda Q}$, the conditioned measure $P_t^{(\lambda)}(C)$ is given by
\begin{subequations}
\begin{equation}
P_t^{(\lambda)} (C)=\frac{ Z^{(\lambda)}_t(C)\, \mathbb{Z}^{(\lambda)}_t(C)}{\sum_{C'}  Z^{(\lambda)}_t(C')\, \mathbb{Z}^{(\lambda)}_t(C')}
\label{eq:cond prob 2 app}
\end{equation}
where $Z^{(\lambda)}_t(C)$ and $\mathbb{Z}^{(\lambda)}_t(C)$ follow the recursion relations
\begin{align}
Z^{(\lambda)}_t(C)=& \sum_{C'}e^{\lambda f_{t-1}(C')+\lambda g_{t-1}(C,C')}M_0(C,C')Z^{(\lambda)}_{t-1}(C') \label{eq:Z 1}\\
\mathbb{Z}^{(\lambda)}_t(C)=& \sum_{C'}e^{\lambda f_{t}(C)+\lambda g_{t}(C',C)}M_0(C',C)\mathbb{Z}^{(\lambda)}_{t+1}(C') \label{eq:Z 2}
\end{align}
\end{subequations}
One can also show that the conditioned dynamics remains Markovian, and $P_t^{(\lambda)} (C)$ follows \eqref{eq:Markov W}
with the transition probability
\begin{align}
W_t^{(\lambda)}(C',C)=&\frac{\mathbb{Z}_{t+1}^{(\lambda)}(C')M_0(C',C)e^{\lambda f_{t}(C)+\lambda g_t(C',C)}Z^{(\lambda)}_t(C)}{\sum_{C''}\mathbb{Z}_{t+1}^{(\lambda)}(C'')M_0(C'',C)e^{\lambda f_{t}(C)+\lambda g_t(C'',C)}Z^{(\lambda)}_t(C)}\cr
=& \frac{\mathbb{Z}_{t+1}^{(\lambda)}(C')}{\mathbb{Z}^{(\lambda)}_t(C)}e^{\lambda f_{t}(C)+\lambda g_t(C',C)}M_0(C',C) \label{eq:W general}
\end{align}
One can verify using \eqref{eq:Z 2} that $\sum_{C'}W_t^{(\lambda)}(C',C)=1$. 

The expressions (\ref{eq:prob I}-\ref{eq:prob V}) and (\ref{eq:tr prob 1}-\ref{eq:tr prob 2}) for $Q=Q_T$ in \eqref{eq:QT} can be easily recovered from \eqref{eq:cond prob 2 app} and \eqref{eq:W general} by using the corresponding $f_t(C)$ and $g_t(C',C)$ and taking large $T$ limit.

\subsection{Continuous time Markov process. \label{sec:continuous time}}
The case of a continuous time Markov process can be obtained by choosing a Markov matrix $M_0$ in the discrete time case of the form
\begin{equation}
M_0(C',C) = \left(1 - \sum_{C^{''}}{\cal M}_0(C^{''},C) dt\right) \delta_{C',C} + {\cal M}_0(C',C) \, dt+\cdots
\label{eq:limit construction}
\end{equation}
and subsequently taking the limit $dt \to 0$ in the corresponding Master equation. The ${\cal M}_0(C',C)$ is the jump rate from configuration $C$ to $C'$. Following this construction it is straightforward to extend the results of conditioned process in the discrete time case to the continuous time. The details are given in Appendix \ref{app:cont time}.

\section{The Langevin dynamics \label{sec:langevin}}
We now extend the above discussion to a Langevin process on the real line defined by the stochastic differential equation
\begin{equation}
\dot{X}_t=F(X_t)+\eta_t
\label{eq:langvin free}
\end{equation}
where $F(x)$ is an external force and $\eta_t$ is a Gaussian white noise of mean zero and covariance $\langle\eta_t\eta_{t'}\rangle=\epsilon\;\delta(t-t')$ with $\epsilon$ being the noise strength.
It is well known \cite{VANKAMPEN2007193} that the probability $P_t(x)$ of the process $X_t$ to be in $x$ at time $t$ follows a Fokker-Planck equation 
\begin{equation}
\frac{d}{dt} P_t(x)=\mathcal{L}_0\cdot P_t(x):=-\frac{d}{dx}\left[F(x)P_t(x)\right]+\frac{\epsilon }{2}\frac{d^2}{dx^2}P_t(x)
\label{eq:FP free}
\end{equation}

\subsection{The tilted Fokker-Planck operator}
Our interest is the dynamics conditioned on an empirical observable
\begin{equation}
Q_T=\int_0^T dt\, f(X_t)+\int_0^T dX_t \; h(X_t)
\label{eq:QT diff}
\end{equation}
where $f$ and $h$ are functions of $X_t$. In writing the second integral we mean a special class of observables whose discrete analogue
\begin{equation}
\int_0^T dX_t \; h(X_t)\equiv \sum_{t}(X_{t+dt}-X_t)\left[\alpha \, h(X_{t+dt})+(1-\alpha)\, h(X_t)\right]
\label{eq:discretization QT}
\end{equation}
with $\alpha\in [0,1]$. The choice $\alpha=0$ corresponds to the \^Ito integral and $\alpha=\frac{1}{2}$ corresponds to the Stratonovich integral in stochastic calculus \cite{MckeanBook}. One may view \eqref{eq:QT diff} as a special case of \eqref{eq:QT}. A large number of relevant empirical observables in statistical physics are of the form \eqref{eq:QT diff}.  For example, integrated current, work, entropy production, empirical density, \textit{etc}  \cite{Lebowitz1999,Jack2010,Jack2015,Chetrite2013,Chetrite2015,TOUCHETTE2017,Maes2008,Maes20082,Lecomte20072,Garrahan2009}.

The Langevin dynamics in \eqref{eq:langvin free} can be viewed as a continuous space and time limit of a jump process on a one-dimensional chain (see \aref{app:derivation FP}). This way, the effective dynamics conditioned on $Q_T$ in \eqref{eq:QT diff} can be obtained from our results in \sref{sec:Markov} by suitably taking the continuous limit. For example, a continuous limit of \eqref{eq:GT recursion} gives (see \aref{app:derivation FP})
\begin{equation}
\frac{d}{dT} G_T^{(\lambda)}(x\vert  y)=\mathcal{L}_\lambda\cdot G_T^{(\lambda)}(x \vert y)
\label{eq:GT recurrence cont}
\end{equation}
where the tilted Fokker-Planck operator \cite{Jack2010,Jack2015,Chetrite2013,Chetrite2015,TOUCHETTE2017,Garrahan2009}
\begin{equation}
\mathcal{L}_\lambda :=\lambda f(x)-\left(\frac{d}{dx}-\lambda h(x)\right)F(x)+\frac{\epsilon}{2}\left(\frac{d}{dx}-\lambda h(x)\right)^2+\epsilon \left(\alpha-\frac{1}{2}\right)\lambda h'(x)
\label{eq:tilted FP}
\end{equation}

For large $T$, one gets, analogous to \eqref{eq:main G},
\begin{equation}
G_{T}^{(\lambda)}(x \vert y)\simeq e^{T\mu(\lambda)}r_{\lambda}(x)\ell_{\lambda}(y)
\label{eq:GT cont}
\end{equation}
where $\mu(\lambda)$ is the largest eigenvalue of $\mathcal{L}_\lambda$ and the corresponding eigenvectors $r_\lambda(x)$ and $\ell_\lambda(x)$ are defined by
\begin{equation}
\mathcal{L}_\lambda \cdot r_{\lambda}(x)=\mu(\lambda)r_{\lambda}(x) \quad\textrm{and}\quad \mathcal{L}^{\dagger}_\lambda \cdot \ell_{\lambda}(x)=\mu(\lambda)\ell_{\lambda}(x)
\label{eq:eigen r l eqn}
\end{equation}
where $\mathcal{L}^{\dagger}_\lambda$ is the operator conjugate to $\mathcal{L}_\lambda$.
\begin{equation}
\mathcal{L}^{\dagger}_\lambda := \lambda f(x)+F(x)\left(\frac{d}{dx}+\lambda h(x)\right)+\frac{\epsilon}{2}\left(\frac{d}{dx}+\lambda h(x)\right)^2+\epsilon \left(\alpha-\frac{1}{2}\right)\lambda h'(x)
\label{eq:conj FP}
\end{equation}

\subsection{Conditioned measure for the Langevin dynamics}

One could similarly derive the conditioned measure and the corresponding rate equation. This way (\ref{eq:prob I}-\ref{eq:prob V}) become, for the continuous analogue $P_t^{(\lambda)}(x)$ of the conditioned measure \eqref{eq:cond prob 1} in the five regions of \fref{fig:fig1} (see the derivation in \aref{app:derivation FP})
\begin{subequations}
\begin{itemize}
\item Region I
\begin{equation}
P_t^{(\lambda)}(x)=\frac{\left[e^{-t \mathcal{L}_0^\dagger}\cdot \ell_\lambda\right](x)\,r_0(x)}{\int dy \,\ell_\lambda(y) \, r_0(y)}
\label{eq:pt Langevin I}
\end{equation}
\item Region II
\begin{equation}
P_t^{(\lambda)}(x)=\frac{\ell_\lambda(x)\left[e^{t\mathcal{L}_\lambda}\cdot r_0\right](x)}{e^{t\mu(\lambda)}\int dy \,\ell_\lambda(y) \, r_0(y)}
\label{eq:pt Langevin II}
\end{equation}
\item Region III
\begin{equation}
P_t^{(\lambda)}(x)=\ell_\lambda(x)r_\lambda(x)
\label{eq:pt Langevin III}
\end{equation}
\item Region IV
\begin{equation}
P_t^{(\lambda)}(x)=\frac{\left[e^{(T-t)\mathcal{L}_\lambda^\dagger}\cdot \ell_0\right](x)\, r_\lambda(x)}{e^{(T-t)\mu(\lambda)}\int dy \, r_\lambda(y)} \qquad \textrm{with}\quad \ell_0(x)=1
\label{eq:pt Langevin IV}
\end{equation}
\item Region V
\begin{equation}
P_t^{(\lambda)}(x)=\frac{\left[e^{(t-T)\mathcal{L}_0}\cdot r_\lambda\right](x)}{\int dy \,r_\lambda(y)} 
\label{eq:pt Langevin V}
\end{equation}
\end{itemize}
\end{subequations}

The time evolution of the conditioned dynamics is described by a Langevin equation \eqref{eq:langvin free} with a modified force $F_t^{(\lambda)}(x)$ which, in general, depends on time. 
The force takes different expressions in the five regions indicated in \fref{fig:fig1}.
\begin{subequations}
\begin{itemize}
\item 
Region I
\begin{equation}
F_t^{(\lambda)}(x)=F(x)+\epsilon\frac{d}{dx}\log\left[e^{-t\mathcal{L}_0^{\dagger}}\cdot\ell_{\lambda}(x) \right]
\label{eq:force I}
\end{equation}
\item Region II and III
\begin{equation}
F_t^{(\lambda)}(x)=F(x)+\epsilon\left(\lambda h(x)+\frac{d}{dx}\log\ell_{\lambda}(x)\right)
\label{eq:force II}
\end{equation}
\item
Region IV
\begin{equation}
F_t^{(\lambda)}(x)=F(x)+\epsilon\left(\lambda h(x)+\frac{d}{dx}\log\left[e^{(T-t)\mathcal{L}_\lambda^{\dagger}}\cdot\ell_0(x) \right]\right)
\end{equation}
\item
Region V
\begin{equation}
F_t^{(\lambda)}(x)=F(x)
\label{eq:force V}
\end{equation}
\end{itemize}
\end{subequations}
A derivation is given in \aref{app:derivation FP}. One can easily verify that the probability (\ref{eq:pt Langevin I}-\ref{eq:pt Langevin V}) follows a Fokker-Planck equation with the corresponding force (\ref{eq:force I}-\ref{eq:force V}). To see this, for example in region I, one can simply use that $\left[e^{-t \mathcal{L}_0^\dagger}\cdot \ell_\lambda\right](x)\equiv V_t(x)$ in \eqref{eq:pt Langevin I} is a solution of $\frac{d}{dt}V_t(x)=-\mathcal{L}_0^\dagger\cdot V_t(x)$ and that $\mathcal{L}_0\cdot r_0(x)=0$.

\begin{remarkk}
We have considered the noise amplitude $\epsilon$ in \eqref{eq:langvin free} to be a constant. A generalization where the amplitude is a function of $X_t$ involves a choice of the \^{I}to-Stratonovich discretization \cite{MckeanBook}. The analysis could be easily extended to such cases as well as in higher dimensions.
\end{remarkk}

\subsection{The Ornstein-Uhlenbeck process \label{sec:example}}
As an illustrative easy example one can consider the Langevin equation in a harmonic potential, $F(x)=-\gamma \, x$. This is known as the Ornstein-Uhlenbeck process \cite{VANKAMPEN2007193}.
To make our discussion simple, we choose the observable $Q_T=\int_0^Tds\; X_s$ which corresponds to $f(x)=x$ and $h(x)=0$ in \eqref{eq:QT diff}. In this case, the tilted Fokker-Planck operator \eqref{eq:tilted FP} gives
\begin{equation*}
\mathcal{L}_\lambda :=\lambda x +\gamma \frac{d}{dx}x+\frac{\epsilon}{2}\frac{d^2}{dx^2}
\label{eq:tilted FP ex}
\end{equation*}
Its largest eigenvalue and the corresponding eigenvectors are
\begin{equation}
\mu(\lambda)=\frac{\epsilon \lambda^2}{2\gamma^2};
\qquad r_\lambda(x)=\mathcal{N}e^{-\frac{\gamma}{\epsilon}\left( x-\frac{\mu}{\lambda}\right)^2}; \qquad \ell_\lambda(x)=e^{\frac{\lambda}{\gamma}x}
\label{eq:mu OU}
\end{equation}
with $\mathcal{N}$ determined from normalization $\int dx \ell_\lambda(x)r_\lambda(x)=1$. The ensemble equivalence \eqref{eq:legendre} gives the large deviations function $\phi(q)=\tfrac{\gamma^2}{2\epsilon}q^2$. 

The conditioned probability (\ref{eq:pt Langevin I}-\ref{eq:pt Langevin V}) and the effective force (\ref{eq:force I}-\ref{eq:force V}) can be explicitly evaluated in this example. One would essentially need to evaluate terms like $\left[e^{-t \mathcal{L}_0^\dagger}\cdot \ell_\lambda\right](x)\equiv V_t(x)$ which is a solution of $\frac{d}{dt}V_t(x)=-\mathcal{L}_0^\dagger\cdot V_t(x)$ with an initial condition $V_0(x)=\ell_\lambda(x)$. It is simple to verify that the solution is
\begin{equation*}
\left[e^{-t \mathcal{L}_0^\dagger}\cdot \ell_\lambda\right](x)=\exp \left[\frac{\lambda x}{\gamma}e^{\gamma t}+\frac{\lambda^2\epsilon }{4\gamma^3}\left(1-e^{2\gamma t}\right)\right] \qquad \textrm{for } t\le 0
\end{equation*} 
Similarly, one can verify
\begin{align*}
&\left[e^{t\mathcal{L}_\lambda}\cdot r_0\right](x)=\mathcal{N}\exp  \left[\left(1-e^{-\gamma t}\right)\left\{\frac{\lambda x}{\gamma}-\frac{\epsilon \lambda^2}{4\gamma^3}\left(3-e^{-\gamma t}\right) \right\}+\frac{\epsilon \lambda^2 t}{2\gamma^2}-\frac{\gamma x^2}{\epsilon}\right]\\
& \qquad  \qquad \qquad \qquad \qquad \qquad \qquad \qquad \qquad \qquad ~~ \qquad \qquad \qquad \textrm{for } t\ge 0, \\
&\left[e^{(T-t)\mathcal{L}_\lambda^\dagger}\cdot \ell_0\right](x)= \exp  \left[\left(1-e^{-\gamma (T-t)}\right)\left\{\frac{\lambda x}{\gamma}-\frac{\epsilon \lambda^2}{4\gamma^3}\left(3-e^{-\gamma (T-t)}\right) \right\}\right.\\
& \qquad \qquad \qquad  \qquad \qquad \qquad \qquad  \qquad \qquad \left.  +\frac{\epsilon \lambda^2 (T-t)}{2\gamma^2}\right]  \qquad \textrm{for } t\le T, \\
&\left[e^{(t-T)\mathcal{L}_0}\cdot r_\lambda\right](x)= \mathcal{N}\exp \left[-\frac{\gamma}{\epsilon}\left(x-\frac{\epsilon \lambda}{2\gamma^2} e^{-\gamma(t-T)}\right)^2\right] \qquad ~ \textrm{for } t\ge T.
\end{align*}

Using these in the general expression (\ref{eq:pt Langevin I}-\ref{eq:pt Langevin V}) and (\ref{eq:force I}-\ref{eq:force V}) we find that, in all regions, the conditioned measure and the effective force are of the form 
\begin{equation}
P_t^{(\lambda)}(x)=\sqrt{\frac{\gamma}{\pi \epsilon }}\exp\left[-\frac{\gamma}{\epsilon}(x-a_t)^2\right]\qquad \textrm{and}\qquad F_t^{(\lambda)}(x)=-\gamma\left(x-\epsilon \, b_t\right)
\label{eq:Pt OU result}
\end{equation}
This means that the conditioned dynamics is another Langevin equation in a harmonic potential whose minimum is at $\epsilon\, b_t$.
We get, in region I, $a_t=\frac{\epsilon \lambda}{2\gamma^2}e^{\gamma t}$ and $b_t=\frac{ \lambda}{\gamma^2}e^{\gamma t}$; in region II,
$a_t=\frac{\epsilon \lambda}{\gamma^2}\left(1-\frac{1}{2}e^{-\gamma \, t}\right)$ and $ b_t=\frac{ \lambda}{\gamma^2}$; in region III, $a_t=\frac{\epsilon \lambda}{\gamma^2}$ and $b_t=\frac{ \lambda}{\gamma^2}$; in region IV, $a_t=\frac{\epsilon \lambda}{\gamma^2}\left(1-\frac{1}{2}e^{-\gamma \, (T-t)}\right)$ and $b_t=\frac{ \lambda}{\gamma^2}\left(1-e^{-\gamma (T-t)}\right)$; in region V, $a_t=\frac{\epsilon \lambda}{2\gamma^2} e^{-\gamma \, (t-T)}$ and $b_t=0$.

\begin{figure}
\centering \includegraphics[width=0.9\textwidth]{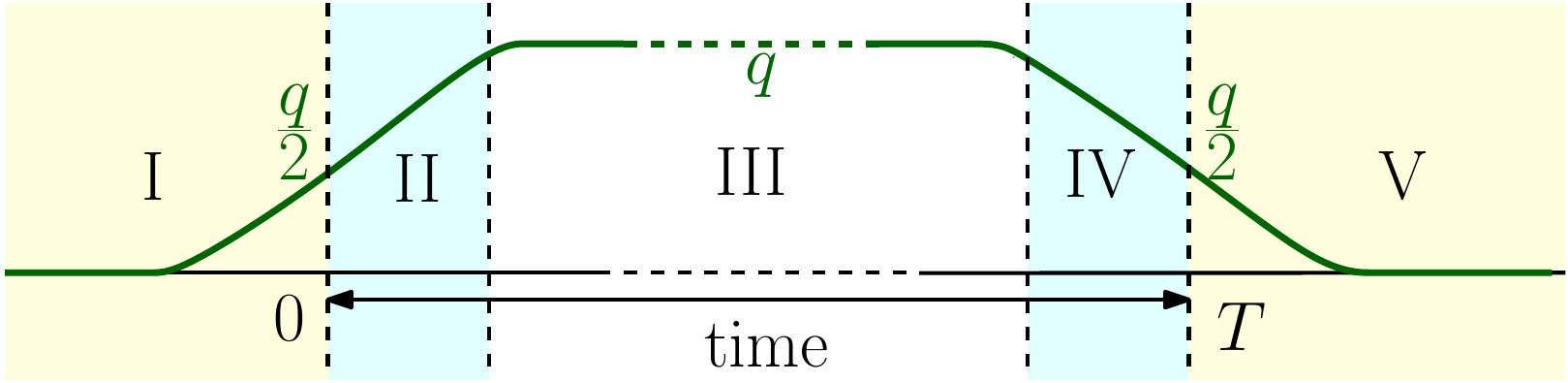}
\caption{A schematic of the most probable trajectory of the conditioned Ornestein-Uhlenbeck process defined in \sref{sec:example}. The most probable position changes with time, only reaching a time independent value $q=\frac{\epsilon \lambda}{\gamma^2}$ at the intermediate quasi-stationary region III. The evolution is symmetric under time reversal, with most probable position $\frac{q}{2}$ at $t=0$ and $t=T$. \label{fig:fig2}}
\end{figure}

One can get the micro-canonical probability $\mathcal{P}_t(x\vert q)$ using $\frac{\epsilon \lambda}{\gamma^2}=q$ in the above expression for $P_t^{(\lambda)}(x)$. From this solution, one can also see that the most likely trajectory followed by the system is $x(t)=a_t$. A schematic of the trajectory is given in \fref{fig:fig2}. 

\begin{remarks}~\\
\begin{enumerate}
\item
In this example, both $X_t$ and $Q_T$ are Gaussian variables. The direct calculation of the covariance could be an alternative way of re-deriving the result \eqref{eq:Pt OU result}.
\item
Here, the conditioned measure $P_t^{(\lambda)}(x)$ is symmetric under $t\rightarrow T-t$, thus symmetric under time reversal. This is because on a one-dimensional line the force $F(x)$ can be written as the gradient of a potential and the Langevin dynamics satisfies detailed balance. This would not necessarily be the case on a ring or in higher dimensions.
\end{enumerate}
\end{remarks}

\section{Large deviations in the conditioned Langevin dynamics. \label{sec:ldf variational big}}
We shall now discuss the Langevin dynamics on the line when the noise strength $\epsilon$ is small. This weak noise limit has been of interest in the past particularly in the Freidlin-Wentzel theory of stochastic differential equations \cite{Freidlin2012}. One may also view the fluctuating hydrodynamics description of interacting many-body systems as a generalization of the Langevin equation where the weak noise limit comes from the large system size \cite{Sadhu2016,Derrida2007,Bertini2014,Bertini2002}. A generalization of our discussion here to a many-body system will be presented in a future publication \cite{Sadhu2018}. 

In this weak noise limit, one can describe rare fluctuations in terms of a large deviations function \cite{Freidlin2012,Graham1985,Graham1973}. For example, the steady state probability of a Langevin equation describing a particle in a potential $U(x)$ has a large deviations form
\begin{equation*}
P(x)\sim e^{-\frac{2}{\epsilon}U(x)}\qquad \textrm{for small $\epsilon$.}
\end{equation*}

In this Section, we shall show that a similar large deviations description holds for the conditioned measure in the Langevin equation.

\subsection{WKB solution of the eigenfunctions \label{sec:deformed ldf}}
For small $\epsilon$, one can try the WKB method \cite{Landau} to determine the largest eigenvalue and associated eigenvectors of the tilted operator $\mathcal{L}_\lambda$ in \eqref{eq:tilted FP}. This means that we look for a solution of the type
\begin{subequations}
\begin{equation}
r_{\frac{\kappa}{\epsilon}}(x)\sim e^{-\frac{1}{\epsilon}\psi_\rt^{(\kappa)}(x)}, \qquad  \ell_{\frac{\kappa}{\epsilon}}(x)\sim e^{-\frac{1}{\epsilon}\psi_\lt^{(\kappa)}(x)}
\label{eq:ldf for r l}
\end{equation}
by setting
\begin{equation}
\lambda=\frac{\kappa}{\epsilon} \qquad \textrm{and}\qquad \mu\left(\frac{\kappa}{\epsilon}\right)\simeq \frac{1}{\epsilon}\chi(\kappa)
\label{eq:wkb lambda}
\end{equation}
\end{subequations}
in the eigenvalue equations \eqref{eq:eigen r l eqn}.
We find that, for small $\epsilon$, this is indeed a consistent solution to the leading order when the large deviations functions satisfy
\begin{subequations}
\begin{align}
F(x)^2-\left(\frac{d}{dx}\psi_\lt^{(\kappa)}(x)-\kappa h(x)-F(x)\right)^2= & 2\kappa f(x)-2\chi(\kappa) \label{eq:HJ1} \\
F(x)^2-\left(\frac{d}{dx}\psi_\rt^{(\kappa)}(x)+\kappa h(x)+F(x)\right)^2= & 2\kappa f(x)-2\chi(\kappa) \label{eq:HJ2}
\end{align}
\end{subequations}
When we use such a solution in \eqref{eq:GT cont} we get
\begin{equation}
G_T^{(\frac{\kappa}{\epsilon})}(x\vert y)\sim e^{\frac{T}{\epsilon}\chi(\kappa)-\frac{1}{\epsilon}\psi_\rt^{(\kappa)}(x)-\frac{1}{\epsilon}\psi_\lt^{(\kappa)}(y)}
\label{eq:G large deviation form}
\end{equation}
for small $\epsilon$. 
This also gives a large deviations form for the conditioned measure. 
In particular, the conditioned measure \eqref{eq:p lambda T} and \eqref{eq:P cond t 0}, for small $\epsilon$, gives
\begin{equation}
\mathcal{P}_T^{(\frac{\kappa}{\epsilon})}(x)\sim e^{-\frac{1}{\epsilon}\psi_T^{(\kappa)}(x)} \qquad \textrm{and} \qquad \mathcal{P}_0^{(\frac{\kappa}{\epsilon})}(x)\sim e^{-\frac{1}{\epsilon}\psi_0^{(\kappa)}(x)}
\label{eq:P at T and 0}
\end{equation}
where $\psi_T^{(\kappa)}(x)\simeq \psi_\rt^{(\kappa)}(x)$ and $\psi_0^{(\kappa)}(x)\simeq\psi_\lt^{(\kappa)}(x)+\mathcal{F}(x)$ up to an additive constant (we denote by $\mathcal{F}(x)$ the large deviations function associated to the steady state probability of the original Langevin equation \eqref{eq:langvin free}).

\begin{remarks}~\\
\begin{enumerate}
\item
The solution \eqref{eq:G large deviation form} implies that the joint probability \eqref{eq:PT large dev} also has a large deviations form given by
\begin{equation*}
P_T(x,Q_T=qT \vert y)\sim e^{-\frac{T}{\epsilon}\Phi(q)-\frac{1}{\epsilon}\psi_\rt(x,q)-\frac{1}{\epsilon}\psi_\lt(y,q)}
\end{equation*}
for small $\epsilon$, and the large deviations functions are related to their counterpart $\chi(\kappa)$, $\psi_\rt^{(\kappa)}(x)$, and $\psi_\lt^{(\kappa)}(x)$ by the ensemble equivalence
\begin{equation}
\Phi(q)=\kappa \; q -\chi(\kappa) \qquad \textrm{with} \qquad \kappa=\Phi'(q)
\label{eq:equivalence small}
\end{equation}
for large $T$.
\item
Later, in \sref{sec:HJ}, we will see that (\ref{eq:HJ1}-\ref{eq:HJ2}) are the Hamilton-Jacobi equations in a variational formulation of the problem. 
\end{enumerate}
\end{remarks}

\subsection{Conditioned large deviations}
The WKB solution \eqref{eq:ldf for r l} gives that the conditioned measure at any time $t$, in the two ensembles, has a large deviations form
\begin{equation}
P_t^{(\frac{\kappa}{\epsilon})}(x)\sim e^{-\frac{1}{\epsilon}\psi_t^{(\kappa)}(x)} \qquad \textrm{and} \qquad \mathcal{P}_t(x\vert Q=qT)\sim e^{-\frac{1}{\epsilon}\psi_t(x,q)}
\label{eq:ldf for Plt}
\end{equation}
with the two conditioned large deviations functions related by the equivalence of ensembles \eqref{eq:equivalence small}. This is already seen in \eqref{eq:P at T and 0}. For other times, this comes from using the WKB solution (\ref{eq:ldf for r l}-\ref{eq:wkb lambda}) in the expressions (\ref{eq:pt Langevin I}-\ref{eq:pt Langevin V}) for small $\epsilon$.

Among these, the simplest case is the quasi-stationary regime, \textit{i.e.} $1\ll t$ and $T-t\gg 1$, where
$P_t^{(\lambda)}(x)=r_\lambda(x)\ell_\lambda(x)$ given in \eqref{eq:pt Langevin III}.  Using \eqref{eq:ldf for r l} we get
\begin{equation}
\psi_t^{(\kappa)}(x)\equiv \psi_\textrm{mid}^{(\kappa)}(x)=\psi_\rt^{(\kappa)}(x)+\psi_\lt^{(\kappa)}(x)
\label{eq:psi m}
\end{equation}
In other regions one could similarly derive expressions for $\psi_t^{(\kappa)}(x)$.

\section{Gradient force \label{sec:gradient}}
For the rest of this paper, we shall consider the Langevin equation \eqref{eq:langvin free} on the line where the 
force is the gradient of a confining potential $U(x)$, \textit{i.e.} $F(x)=-\partial_x U(x)$.  For simplicity we shall only consider $Q_T=\int_0^Tdt\, f(X_t)$ (\textit{i.e.} $h(x)=0$ in \eqref{eq:QT diff}). 

As a consequence, the two solutions of the Hamilton-Jacobi equations (\ref{eq:HJ1}-\ref{eq:HJ2}) are related,
\begin{equation}
\psi_\lt^{(\kappa)}(x)=\psi_\rt^{(\kappa)}(x)-2U(x)+\textrm{constant}
\label{eq:symmetry psi l r}
\end{equation}
(This would not be true, in general, when $F(x)$ is not a gradient of a potential. For example, on a ring with a circular driving force.)

 Moreover, using \eqref{eq:symmetry psi l r}, the effective force \eqref{eq:force II} in the quasi-stationary regime, for small $\epsilon$, can be written as
\begin{equation}
F_t^{(\frac{\kappa}{\epsilon})}(x)\simeq F(x)-\frac{d}{dx}\psi_\lt^{(\kappa)}(x)=-\frac{1}{2}\partial_x \psi_\textrm{mid}^{(\kappa)}(x)
\label{eq:effective force ldf}
\end{equation}
(This is only the leading order term for small $\epsilon$.) This shows that the conditioned process can be viewed as a Langevin dynamics in the potential landscape of the conditioned large deviations function.

\subsubsection*{An explicit solution \label{sec:cond Langevin special}}
The Hamilton-Jacobi equations (\ref{eq:HJ1}-\ref{eq:HJ2}) are simple to solve. 
For example, lets take 
\eqref{eq:HJ2} which is quadratic and has two solutions $\psi_\pm^{(\kappa)}(x)$ which follows
\begin{equation*}
\partial_x\psi_\pm^{(\kappa)}(x)=-F(x)\pm \sqrt{F(x)^2-2\kappa f(x)+2\chi(\kappa)}
\end{equation*}
When $F(x)^2-2\kappa f(x)$ has a single global minimum at a value $x=u$ and it grows at $x\rightarrow\pm \infty$ (and $F(x)$ is a gradient of a confining potential), the only possible choice is that
\begin{equation*}
\partial_x\psi_\rt^{(\kappa)}(x)=\begin{cases}\partial_x\psi_{+}^{(\kappa)}(x), & \mbox{for } x\ge u, \\ \partial_x\psi_{-}^{(\kappa)}(x), & \mbox{for } x\le u. \end{cases}
\end{equation*}
At the meeting point, the eigenfunction $r_{\frac{\kappa}{\epsilon}}(x)$ and its derivative are continuous which leads to continuity of $\partial_x\psi_\rt^{(\kappa)}(x)$. The latter condition gives
\begin{equation}
\chi(\kappa)=\kappa f(u)-\frac{1}{2}F(u)^2 \qquad \textrm{with}\qquad \kappa=\frac{F(u)F'(u)}{f'(u)}
\label{eq:chi kappa expr}
\end{equation}

\begin{remarkk} The reason for imposing the condition that $F(x)^2-2\kappa f(x)$ has a single global minimum is that otherwise, one can not straightforwardly extend the asymptotic solutions $\psi_\pm^{(\kappa)}(x)$ to all values of $x$, similar to the WKB analysis of double well potential in Quantum Mechanics \cite{Landau}. This is because between the minima the eigenfunction is a superposition of the $\psi_{+}^{(\kappa)}(x)$ and $\psi_{-}^{(\kappa)}(x)$ solutions and one has to carefully match the solutions at each minimum. \end{remarkk}

The second Hamilton-Jacobi equation \eqref{eq:HJ1} is similarly solved. Integrating these solutions we write 
\begin{subequations}
\begin{align}
\psi_\rt^{(\kappa)}(x)=&\int_{x^\star}^{x}dz\;\left\{-F(z)+\textrm{sgn}(x-u)\sqrt{F(z)^2-F(u)^2-2\kappa[f(z)-f(u)]}\right\} \label{eq:psiT sol}\\
\psi_\lt^{(\kappa)}(x)=&K+\int_{x^\star}^{x}dz\;\left\{F(z)+\textrm{sgn}(x-u)\sqrt{F(z)^2-F(u)^2-2\kappa[f(z)-f(u)]}\right\} \label{eq:psi0 sol}
\end{align}
\end{subequations}
where $K$ and $x^\star$ are a priori arbitrary constants. To satisfy the normalization $\int dx \, r_\lambda(x)\ell_\lambda(x)=1$, one can choose $K=0$ for $x^\star=u$ (using $F(x)^2-2\kappa f(x)$ has minimum at $x=u$).

Using (\ref{eq:psiT sol}-\ref{eq:psi0 sol}) in \eqref{eq:P at T and 0} one can see that $\psi_T^{(\kappa)}(x)$ and $\psi_0^{(\kappa)}(x)$ both have minimum at $x_0$ given by $f(x_0)=f(u)-\frac{1}{2\kappa}F(u)^2$. This makes $x_0$ the most likely position at time $t=0$ and $t=T$ which is different from the quasi-stationary position $u$.

As a consequence of (\ref{eq:psiT sol}-\ref{eq:psi0 sol}) we get the conditioned large deviations function \eqref{eq:psi m} in the quasi-stationary regime  
\begin{equation}
\psi_\textrm{mid}^{(\kappa)}(x)=2\,\textrm{sgn}(x-u)\int_{u}^{x}dz\;\sqrt{F(z)^2-F(u)^2-2\kappa[f(z)-f(u)]}
\label{eq:psi m example}
\end{equation}
This shows that $x=u$ is the most likely position in the quasi-stationary regime.

\begin{remarks}~\\
\begin{enumerate}
\item
In this example, one could systematically calculate sub-leading corrections in the eigenvalue and eigenvector. Writing
\begin{equation*}
r_{\frac{\kappa}{\epsilon}}(x)= e^{-\frac{1}{\epsilon}\psi_\rt^{(\kappa)}(x)-\widetilde{\psi}_\rt^{(\kappa)}(x)+\cdots}, \qquad \mu\left(\frac{\kappa}{\epsilon}\right)= \frac{1}{\epsilon}\chi(\kappa)+\widetilde{\chi}(\kappa)+\cdots
\end{equation*}
in \eqref{eq:eigen r l eqn} (we are using $h(x)=0$) and expanding in powers of $\epsilon$ one would get in the sub-leading order 
\begin{equation}
-F'(x)+\left[F(x)+\partial_x\psi_\rt^{(\kappa)}(x)\right]\partial_x \widetilde{\psi}_\rt^{(\kappa)}(x)-\frac{1}{2}\partial_x^2\psi_\rt^{(\kappa)}(x)=\widetilde{\chi}(\kappa)
\label{eq:chi tilde}
\end{equation}
Using \eqref{eq:psiT sol} we see that the term $F(x)+\partial_x\psi_\rt^{(\kappa)}(x)$ in \eqref{eq:chi tilde} vanishes at $x=u$.  Moreover, from \eqref{eq:psiT sol} we get
\begin{equation*}
\lim_{x\rightarrow u}\partial_x^2\psi_\rt^{(\kappa)}(x)=-F'(u)+\sqrt{F'(u)^2+F(u)F''(u)-\kappa f''(u)}
\end{equation*}
This and the fact that $\partial_x\psi_\rt^{(\kappa)}(x)=-F(x)$ for $x=u$ gives for the sub-leading order correction to the eigenvalue 
\begin{equation}
\widetilde{\chi}(\kappa)=-\frac{1}{2}\left[ F'(u)+\sqrt{F'(u)^2+F(u)F''(u)-\kappa f''(u)}\right]
\end{equation}
An explicit expression for $\widetilde{\psi}_\rt^{(\kappa)}(x)$ could also be deduced from \eqref{eq:psiT sol} and \eqref{eq:chi tilde}.
\item
One can also check that the results for the Ornstein-Uhlenbeck process in \sref{sec:example} can be recovered by choosing $f(x)=x$ and $F(x)=-\gamma x$.
\end{enumerate}
\end{remarks}

\section{A variational formulation \label{sec:variational}}
The path integral formulation of the Langevin equation offers an alternative approach for the conditioned dynamics. In this, the conditioned large deviations function is obtained as a solution of a variational problem. As in \sref{sec:gradient}, we consider a gradient force and $Q_T=\int_0^Tdt\, f(X_t)$, although one could extend the analysis for other cases.

We introduce the formulation for the generating function $G_T^{(\lambda)}(x\vert y)$ for the Langevin dynamics. Using a path integral solution of \eqref{eq:GT recurrence cont} (see \aref{app:Fok to Path} for details) one can write, for small $\epsilon$,
\begin{equation}
G_T^{(\frac{\kappa}{\epsilon})}(x\vert y)\sim \int_{z(0)=y}^{z(T)=x}\mathcal{D}[z]e^{\frac{1}{\epsilon}S_T^{(\kappa)}[z(t)]}
\label{eq:G path small}
\end{equation}
where the Action 
\begin{equation}
S_T^{(\kappa)}[z]=\int_0^Tdt\left\{\kappa\, f(z)-\frac{\dot{z}^2}{2}+ \dot{z}\, F(z)-\frac{F(z)^2}{2} \right\}
\label{eq:sT}
\end{equation}
One may view \eqref{eq:G path small} as a sum over all paths (connecting $y$ to $x$ during time $T$) weighted by $\exp(\frac{1}{\epsilon}S_T^{(\kappa)}[z])$.

In the small $\epsilon$ limit, if we assume that \eqref{eq:G path small} is dominated by a single path, we get \eqref{eq:G large deviation form} with
\begin{equation}
T\chi(\kappa)-\psi_\rt^{(\kappa)}(x)-\psi_\lt^{(\kappa)}(y)= \max_{z(t)}S_T^{(\kappa)}[z(t)]
\label{eq:saddle action}
\end{equation}
where the maximum is over all possible trajectories $z(t)$ with $z(0)=y$ and $z(T)=x$.

\subsection{An explicit solution \label{sec:example variational}}
Let us first show how this variational approach allows one to recover the results of \sref{sec:cond Langevin special}. As before, we limit our discussions to the case where  $F^2(z)-2\kappa \, f(z)$ has a single global minimum at $x=u$. It will be clear shortly, that in the variational formulation, this condition ensures a single time independent optimal path.
\begin{figure}
\centering \includegraphics[width=0.9\textwidth]{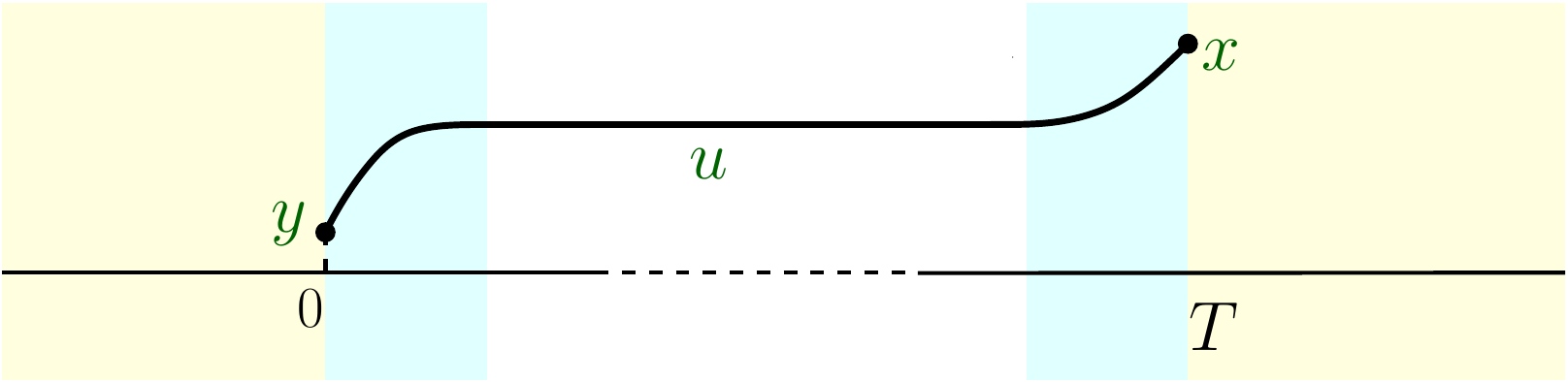}
\caption{A schematic of the optimal path for the variational problem in \sref{sec:example variational}. \label{fig:fig2.5}}
\end{figure}

Using variational calculus we get from (\ref{eq:G path small}-\ref{eq:sT}) that the optimal path follows
\begin{equation*}
\ddot{z}=\frac{d}{dz}\left[\frac{F(z)^2}{2}-\kappa f(z)\right]
\end{equation*}
Multiplying the above equation with $2 \dot{z}$ and integrating we get
\begin{equation*}
\dot{z}^2=F(z)^2-2\kappa f(z)+K
\end{equation*}
where $K$ is an integration constant. We see the similarity with the trajectory of a mechanical particle of constant energy $\frac{1}{2}K$ in a potential $\kappa f(z)-\frac{F(z)^2}{2}$ which has a single global maximum at $x=u$. The trajectory has to cover a finite distance from the point $y$ to the point $x$ in a very large time $T$. The only possible way this could happen if the trajectory passes arbitrarily close to $u$ which is a repulsive fixed point of the mechanical dynamics. This requires an energy almost equal to the maximum of the mechanical potential, with the difference vanishing as $T$ grows. This gives $K=2\kappa f(u)-F(u)^2$ and the optimal path
\begin{equation}
\dot{z}^2=F(z)^2-2\kappa f(z)+2 \kappa f(u)-F(u)^2
\label{eq:dot z sqr}
\end{equation}

Such a trajectory spends most of its time in the position $u$, and deviates from it only near the boundary to comply with the condition $z(0)=y$ and $z(T)=x$, as sketched in \fref{fig:fig2.5}. Then, we can write the optimal path \eqref{eq:dot z sqr}, for large $T$, as
\begin{equation*}
\dot{z}(t)=\begin{cases}\textrm{sgn}(u-y)\sqrt{F(z)^2-2\kappa f(z)+2 \kappa f(u)-F(u)^2}, & \textrm{for $0\le t\ll T$,} \\0, & \textrm{for $1\ll t$ and $T-t\gg 1$,}\\ \textrm{sgn}(x-u)\sqrt{F(z)^2-2\kappa f(z)+2 \kappa f(u)-F(u)^2}, & \textrm{for $0\le T-t\ll T$.}
\end{cases}
\end{equation*}

To use this in the variational formula \eqref{eq:saddle action} we substitute $F(z)^2$ from \eqref{eq:dot z sqr} in the expression \eqref{eq:sT} and get
\begin{equation*}
\max_{z(t)}S_T^{(\kappa)}[z(t)]=T \left[ \kappa \, f(u)-\frac{1}{2}F(u)^2\right]+\int_0^{t_0} dt\, \dot{z}\, \left[F(z)-\dot{z}\right]+\int_{t_0}^T dt\, \dot{z}\left[F(z)-\dot{z}\right]
\end{equation*}
where $t_0\in [0,T]$.
We see that, the integration variable can be changed to $z$, and when $1\ll t_0$ and $T-t_0\gg 1$, we can use $z(t_0)=u$, in addition to the boundary condition $z(0)=y$ and $z(T)=x$. Using the explicit solution of $\dot{z}(t)$, given above, we get
\begin{align*}
\max_{z(t)}S_T^{(\kappa)}&[z(t)]=T \left[ \kappa \, f(u)-\frac{1}{2}F(u)^2\right]\\
-&\int_u^{y} dz\, \left[F(z)+\textrm{sgn}(y-u)\sqrt{F(z)^2-2\kappa f(z)+2 \kappa f(u)-F(u)^2}\right]\\
-&\int_u^{x} dz\, \left[-F(z)+\textrm{sgn}(x-u)\sqrt{F(z)^2-2\kappa f(z)+2 \kappa f(u)-F(u)^2}\right]
\end{align*}
When we use this result in the variational formula \eqref{eq:saddle action} for large $T$, we get $\chi(\kappa)=\left[ \kappa \, f(u)-\frac{1}{2}F(u)^2\right]$, in agreement with our earlier result in \eqref{eq:chi kappa expr}. Moreover, we see that the second and third term gives $\psi_\lt^{(\kappa)}(y)$ and $\psi_\rt^{(\kappa)}(x)$ in (\ref{eq:psiT sol}-\ref{eq:psi0 sol}).


\subsection{Conditioned large deviations function}
One could write a similar variational formula for the conditioned large deviations function $\psi_t^{(\kappa)}(x)$ at an arbitrary time $t$. For large $T$,
\begin{subequations}
\begin{equation}
\psi_t^{(\kappa)}(x)\simeq \max_{z}A_T^{(\kappa)}[z(\tau)]- \max_{z(t)=x}A_T^{(\kappa)}[z(\tau)]
\label{eq:variational 2}
\end{equation}
where the action 
\begin{equation}
A_T^{(\kappa)}[z(\tau)]=\int_{-\infty}^{\infty}d\tau\left\{a(\tau)\, f(z)-\frac{\dot{z}^2}{2}+ \dot{z}\, F(z)-\frac{F(z)^2}{2} \right\}
\label{eq:Action general}
\end{equation}
\end{subequations}
with $a(\tau)=\kappa$ for $\tau\in [0,T]$ and $a(\tau)=0$ elsewhere. The first maximization in \eqref{eq:variational 2} is over all paths, whereas the second maximization is over paths which are conditioned to be at $z(\tau)=x$ for $\tau=t$. 

One may understand the formula \eqref{eq:variational 2} as an optimal contribution from an ensemble of paths with probability weight $e^{\frac{1}{\epsilon} A_T^{(\kappa)}[z]}$ conditioned to pass through $x$ at time $t$; the first term in \eqref{eq:variational 2} is due to normalization.

Here, we show how one can use this variational approach to derive the conditioned large deviations function at an arbitrary time. For this we impose as in \sref{sec:example variational} that $F(x)^2-2\kappa \, f(x)$ has a single global minimum such that the most likely position in the quasi-stationary regime is time independent, $z(\tau)=u$.

\subsubsection*{Quasi-stationary regime.}
\begin{figure}
\centering \includegraphics[width=0.9\textwidth]{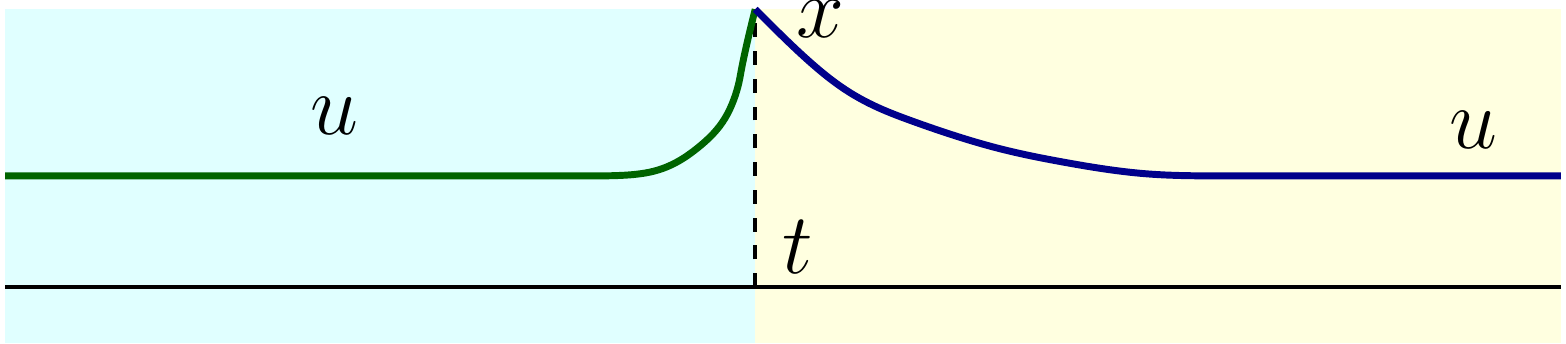}
\caption{A schematic of a path leading to a fluctuation $x$ at time $t$, and subsequent relaxation to the quasi-stationary value $u$ in region III. \label{fig:fig3}}
\end{figure}

Among all the five regions in \fref{fig:fig1}, the simplest is to analyze the quasi-stationary regime where $1\ll t$ and $T-t\gg 1$. Here, for the optimization in \eqref{eq:variational 2}, one essentially need to consider paths which asymptotically reach $u$, both at small $t$, as well as when $t$ is close to $T$. A schematic such path is given in \fref{fig:fig3}.
The analysis is quite similar to that in \sref{sec:example variational}. We get that the optimal path follows
\begin{equation}
\frac{dz(\tau)}{d\tau}=\begin{cases}\textrm{sgn}(x-u)\sqrt{F(z)^2-2\kappa f(z)+2 \kappa f(u)-F(u)^2}, & \textrm{for $\tau <t$}\\ \textrm{sgn}(u-x)\sqrt{F(z)^2-2\kappa f(z)+2 \kappa f(u)-F(u)^2}, & \textrm{for $ \tau>t$,}
\end{cases}
\label{eq:explicit z III}
\end{equation}
and using this in \eqref{eq:variational 2} we get
\begin{equation*}
\psi_t^{(\kappa)}(x)=\int_0^t d\tau\, \dot{z} \left[\dot{z}-F(z)\right]+\int_t^T d\tau\, \dot{z} \left[\dot{z}-F(z)\right]
\end{equation*}
Changing the integration variable to $z$ and using the solution \eqref{eq:explicit z III} with the asymptotics sketched in \fref{fig:fig3}, we get
\begin{align*}
\psi_t^{(\kappa)}(x)=&\int_u^x dz \left[-F(z)+\textrm{sgn}(x-u)\sqrt{F(z)^2-2\kappa f(z)+2 \kappa f(u)-F(u)^2}\right]\\
&+\int_u^x dz \left[F(z)+\textrm{sgn}(x-u)\sqrt{F(z)^2-2\kappa f(z)+2 \kappa f(u)-F(u)^2}\right]
\end{align*}
Comparing with the expression in (\ref{eq:psiT sol}-\ref{eq:psi0 sol}) we see that $\psi_t^{(\kappa)}(x)=\psi_\rt^{(\kappa)}(x)+\psi_\lt^{(\kappa)}(x)$, in agreement with our earlier result \eqref{eq:psi m} and \eqref{eq:psi m example}.

\begin{remarkk}
From \eqref{eq:explicit z III} one could see that the optimal path leading to a fluctuation in the quasi-stationary regime and subsequent relaxation follows a deterministic evolution in a potential landscape of conditioned large deviations function.
\begin{subequations}
\begin{align}
\frac{dz(\tau)}{d\tau}= & F(z)+\frac{d}{dz} \psi_\rt^{(\kappa)}(z)=-\frac{d}{dz}\left[U(z)- \psi_\rt^{(\kappa)}(z)\right] \quad \mbox{for } \tau<t, \label{eq:effective dynamics ldf a} \\ \frac{dz(\tau)}{d\tau}= & F(z)-\frac{d}{dz} \psi_\lt^{(\kappa)}(z)=-\frac{d}{dz}\left[U(z)+\psi_\lt^{(\kappa)}(z)\right] \qquad \mbox{for } \tau>t. \label{eq:effective dynamics ldf b}
\end{align}
\end{subequations}
\end{remarkk}

\subsubsection*{Region II $(0\le t \ll T)$.}
\begin{figure}
\centering \includegraphics[width=0.9\textwidth]{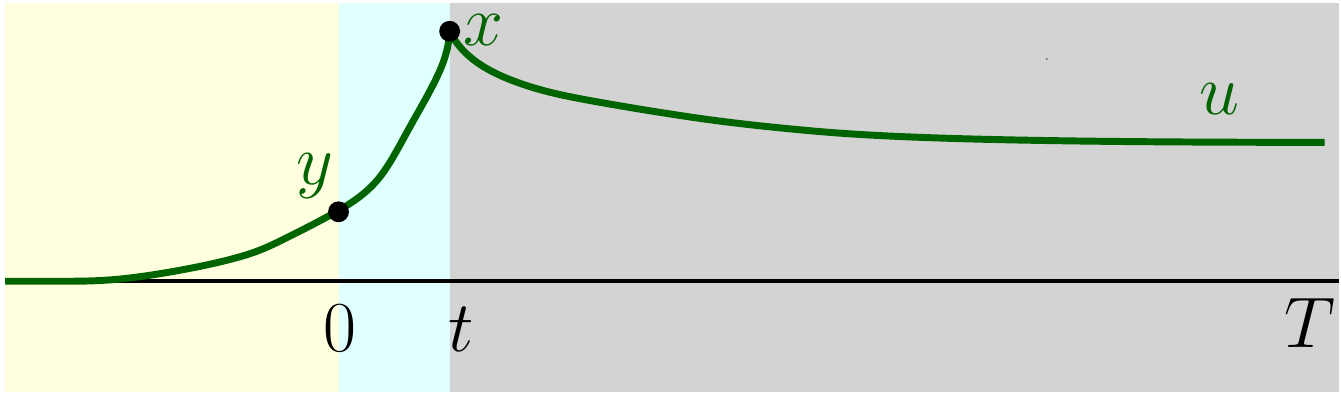}
\caption{A schematic of a path leading to a fluctuation $x$ at $t$ in region II, and subsequent relaxation to the quasistationary position $u$. \label{fig:fig5}}
\end{figure}

The calculation of $\psi_t^{(\kappa)}(x)$ in other regions of time is quite similar. For example, in region II, in the variational formula  \eqref{eq:variational 2}, one essentially need to consider paths which started at the minimum of $U(x)$ (with $F(x)=-U'(x)$) when $\tau\rightarrow -\infty$, pass through $z=x$ at $\tau=t\ge 0$, and asymptotically reach the quasi-stationary value $u$ for large time $\tau\gg 1$, as illustrated in \fref{fig:fig5}.

Following an analysis similar to that in \sref{sec:example variational} it is straightforward to show that the optimal path in this case
\begin{equation}
\dot{z}(\tau)=\begin{cases}-F(z), & \textrm{for $\tau\le 0$}\\\textrm{sgn}(x-y)\sqrt{F(z)^2-2\kappa f(z)+K_1}, & \textrm{for $0\le \tau\le t$}\\ \textrm{sgn}(u-x)\sqrt{F(z)^2-2\kappa f(z)+K_2}, & \textrm{for $\tau \ge t$,}
\end{cases}
\label{eq:sol z II}
\end{equation}
where $K_1$ and $K_2$ are integration constants, and the optimal path passes through $z(0)=y$ (say) when $\tau=0$. The solution for $\tau\le 0$ is easy to see from the condition that at $\tau\rightarrow -\infty$ the system started at the minimum of the potential $U(z)$ with $F(z)=-U'(z)$. Similar asymptotics that for large time the system relaxes to the quasi-stationary position $z=u$ gives the constant $K_2=2 \kappa f(u)-F(u)^2$. In addition, we have the condition
\begin{equation}
t=\int_0^td\tau=\int_y^{x}\frac{dz}{\dot{z}}=\int_y^{x}\frac{dz}{\textrm{sgn}(x-y)\sqrt{F(z)^2-2\kappa f(z)+K_1}}
\label{eq:cond t}
\end{equation}
where we used the solution \eqref{eq:sol z II} and this fixes the constant $K_1$. 

When we use the solution \eqref{eq:sol z II} to write $F(z)^2$ in the expression \eqref{eq:Action general}, we get
\begin{equation*}
\max_{z(t)=x}A_T^{(\kappa)}[z(\tau)]=(T-t)\left[ \kappa f(u)-\frac{F(u)^2}{2}\right]+t\,\frac{K_1}{2}-\int_{-\infty}^T d\tau\,\dot{z}\big[\dot{z}-F(z)\big]
\end{equation*}
Using this in \eqref{eq:variational 2} and the result that $\max_{z(t)}A_T^{(\kappa)}[z(\tau)]=T\left[ \kappa f(u)-\frac{F(u)^2}{2}\right]$, we get
 \begin{equation*}
\psi_t^{(\kappa)}(x)=t \left[ \kappa f(u)-\frac{F(u)^2}{2}\right]-t\,\frac{K_1}{2}+\int_{-\infty}^T d\tau\,\dot{z}\big[\dot{z}-F(z)\big]
\end{equation*}
In this expression, the integration variable can be changed from $\tau$ to $z$, and then using the explicit solution \eqref{eq:sol z II}, we get
\begin{subequations}
\begin{equation}
\psi_t^{(\kappa)}(x)=t \left[ \kappa f(u)-\frac{F(u)^2}{2}\right]+\psi_\lt^{(\kappa)}(x)+\widehat{B}_t^{(\kappa)}(x,y)+\mathcal{F}(y)
\label{eq:app psi II}
\end{equation}
where $\psi_\lt^{(\kappa)}(x)$ is given in \eqref{eq:psi0 sol}, $\mathcal{F}(y)=-2\int_0^y dz\, F(z)$ and
\begin{equation}
\widehat{B}_t^{(\kappa)}(x,y)=-t\,\frac{K_1}{2}+\int_y^x dz\left[-F(z)+\textrm{sgn}(x-y)\sqrt{F(z)^2-2\kappa f(z)+K_1} \right]
\label{eq:B solution}
\end{equation}
\end{subequations}

We note that the condition \eqref{eq:cond t} is equivalent to $\partial_{K_1}\widehat{B}_t^{(\kappa)}(x,y)=0$, which relates $K_1$ to $y$. In addition, the solution \eqref{eq:app psi II} must be optimal over a variation in $y$. These two conditions together leads to $\partial_y\widehat{B}_t^{(\kappa)}(x,y)=2F(y)$, which with the formula \eqref{eq:B solution} gives $K_1=2\kappa\, f(y)$. We note that this is equivalent of continuity of $\dot{z}(\tau)$ at $\tau=0$ in the solution \eqref{eq:sol z II}. This result for $K_1$, along with \eqref{eq:cond t} and (\ref{eq:app psi II}-\ref{eq:B solution}) gives a parametric solution of $\psi_t^{(\kappa)}(x)$ in region II.

We have checked that the same result could be derived using the eigenfunction of the tilted Fokker-Planck operator discussed earlier in \sref{sec:ldf variational big}. 

\subsection{The Hamilton-Jacobi equations from the variational approach	\label{sec:HJ}}
In \sref{sec:deformed ldf} we have shown how one can write the conditioned large deviations function in terms of a solution of the Hamilton-Jacobi equations (\ref{eq:HJ1},\,\ref{eq:HJ2}) derived from the tilted Fokker-Planck operator. In this section, we describe how the same equations can be obtained using the variational formulation in \sref{sec:variational}. The advantage is that in more general problems, \textit{e.g.} the fluctuating hydrodynamics of interacting many-body systems, this variational approach is simpler than using the tilted Fokker-Planck operator (see our future publication \cite{Sadhu2018}).

\begin{figure}
\centering \includegraphics[width=0.9\textwidth]{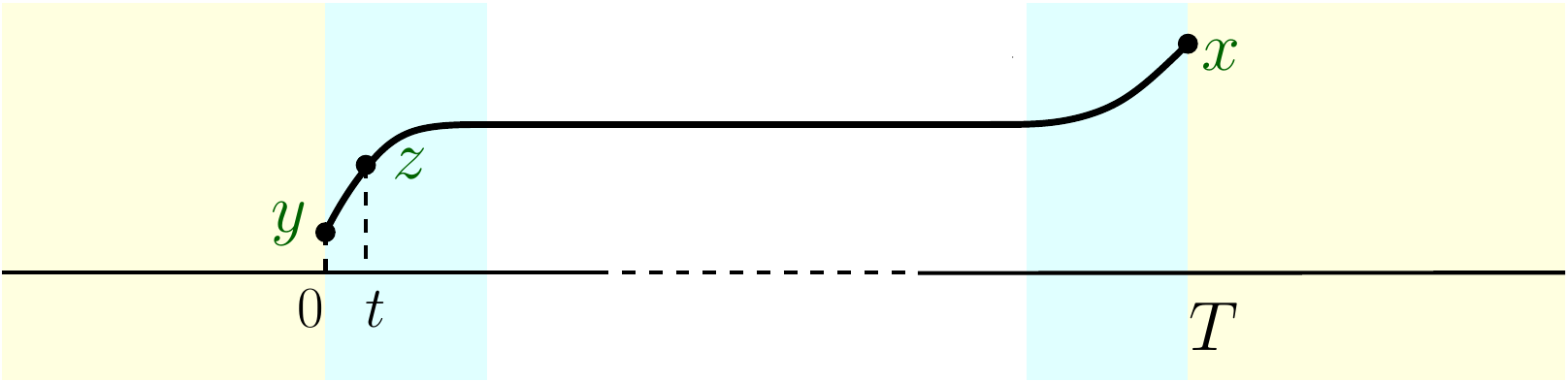}
\caption{A schematic of the sample of paths contributing in the time convolution in \eqref{eq:GG convoluction}. \label{fig:fig4}}
\end{figure}

We start with a derivation of \eqref{eq:HJ1}. Using the definition \eqref{eq:G} one can write for the Langevin equation
\begin{equation}
G_T^{(\lambda)}(x\vert y)=\int dz\, G_{T-t}^{(\lambda)}(x\vert z)G_{t}^{(\lambda)}(z\vert y)
\label{eq:GG convoluction}
\end{equation}
A schematic illustrating this time convolution is shown in \fref{fig:fig4}.
Using the large deviations form \eqref{eq:G large deviation form} and the path integral representation \eqref{eq:G path small}, for small $\epsilon$, it is straightforward to write
\begin{equation}
t\,\chi(\kappa)-\psi_\lt^{(\kappa)}(y)\simeq \max_z\left\{S_{t}^{(\kappa)}(z, y)-\psi_\lt^{(\kappa)}(z)\right\}
\label{eq:chi some int}
\end{equation}
where, from the Action \eqref{eq:sT}, we get for small $t$,
\begin{equation*}
S_{t}^{(\kappa)}(z, y)= t \, \kappa f(y)-\frac{1}{2}\left[ \frac{(z-y)}{t}-F(y)\right]^2+\cdots
\end{equation*}
Expanding \eqref{eq:chi some int} around $y$ we get
\begin{equation*}
\chi(\kappa)\simeq \kappa\, f(y)-\frac{F(y)^2}{2}+\frac{1}{t}\max_z\left\{(z-y)[F(y)-\partial_y\psi_\lt^{(\kappa)}(y)]-\frac{( z-y)^2}{2\,t}\right\}
\end{equation*}
Higher order terms in the expansion are negligible in the small $t$ limit.

In this expression, the maximum is for 
\begin{equation*}
\frac{( z-y)}{t}=F(y)-\partial_y\psi_\lt^{(\kappa)}(y)
\end{equation*}
Substituting this in the above expression for $\chi(\kappa)$ and taking $t\rightarrow 0$ limit we recover the Hamilton-Jacobi equation \eqref{eq:HJ1} for $h(x)=0$.
One can similarly derive the Hamilton-Jacobi equation \eqref{eq:HJ2} for $\psi_\rt^{(\kappa)}(x)$. The analysis could be extended for $h(x)\ne 0$, as well.

\section{The effect of conditioning on the noise. \label{sec:effective noise}}
In (\ref{eq:force I}-\ref{eq:force V}) we have seen that the Langevin dynamics conditioned on $Q_T=\int_0^Tdt\, f(x)$ can be described by another Langevin dynamics with an effective force $F_t^{(\lambda)}(x)$ and white noise $\widetilde{\eta}_t$. In the weak noise limit,  the effective force in the quasi-stationary regime ($t\gg 1$ and $T-t\gg 1$) is given by \eqref{eq:effective force ldf} with \eqref{eq:psi m example}. So the conditioned dynamics, for small $\epsilon$, is
\begin{equation*}
\dot{X}_t=-\textrm{sgn}(X_t-u)\sqrt{F(X_t)^2-F(u)^2-2\kappa[f(X_t)-f(u)]}+\widetilde{\eta}_t
\end{equation*}
where $\widetilde{\eta}_t$ is a Gaussian white noise as in the original (unconditioned) Langevin equation \eqref{eq:langvin free}.

In this quasi-stationary regime, the most probable position $X_t=u$ is time independent (under the condition that $F(x)^2-2\kappa f(x)$ has a single global minimum). Writing small fluctuations $r_t=X_t-u$ we get from the above equation
\begin{equation*}
\dot{r}_t=-\Gamma_u \, r_t+\widetilde{\eta}_t, \qquad \textrm{with}\qquad \Gamma_u=\sqrt{F'(u)^2+F(u)F''(u)-\kappa f''(u)}
\end{equation*}
The solution
\begin{equation*}
r_t=\int_{-\infty}^t ds\, e^{-\Gamma_u\, (t-s)}\,\widetilde{\eta}_s
\end{equation*}
leads to the following correlation
\begin{equation}
\llangle X_tX_{t'}\rrangle_c=\llangle r_t r_{t'}\rrangle=\frac{\epsilon}{2\Gamma_u}e^{-\Gamma_u\vert t-t'\vert}
\label{eq:two time corr}
\end{equation}

If we come back to the original Langevin equation \eqref{eq:langvin free},
\begin{equation}
\dot{X}_t=F(X_t)+\eta_t
\label{eq:non Markovian}
\end{equation}
then the original noise, when the fluctuation $r_t=X_t-u$ is small, is given by
\begin{equation}
\eta_t\simeq -F(u)+\dot{r}_t-F'(u)\, r_t
\label{eq:eta tilde}
\end{equation}
Then, using the correlation \eqref{eq:two time corr} one gets
\begin{subequations}
\begin{equation}
\llangle X_t \, \eta_{t'}\rrangle_c=\llangle r_t\, \eta_{t'} \rrangle=\begin{cases}g_R(t'-t), & \textrm{ for $t'> t$}\\ g_F(t-t'), & \textrm{ for $t'< t$} \end{cases}
\end{equation}
where
\begin{equation}
g_F(t)={ - F'(u)+\Gamma_u  \over 2 \Gamma_u}e^{-\Gamma_u t}\qquad \textrm{and}\qquad g_R(t)={ -F'(u)- \Gamma_u  \over 2 \Gamma_u}e^{-\Gamma_u t}
\end{equation}
\end{subequations}
In this description, we see that the fluctuation $r_t$ is correlated not only to the noise in the past, but also to the noise in the future.
Of course, when one removes the conditioning, \textit{i.e.} for $\kappa=0$, and as a result $u=0$, one has $\Gamma_0=F'(0)$ and $g_R=0$, as one would expect in a Markovian process. One can also show, using \eqref{eq:two time corr} and \eqref{eq:eta tilde}, that
\begin{equation}
\llangle \eta_t \rrangle =-F(u)\qquad \textrm{and}\qquad \langle \eta_t\eta_{t'}\rangle=\epsilon\, \frac{F'(u)^2-\Gamma_u^2}{2\Gamma_u}\, e^{-\Gamma_u\vert t-t'\vert}
\label{eq:new noise correlation}
\end{equation}
so that the original noise $\eta_t$ becomes colored due to the conditioning.

\section{Summary}
In this work we have seen how a stochastic system adapts its dynamics when it is conditioned on a certain value of an empirical observable $Q_T$ of the form \eqref{eq:QT}. The constrained dynamics remains Markovian (see (\ref{eq:Markov W},\,\ref{eq:W general})) if the original process is itself Markovian. In the case of the Langevin dynamics, the conditioning modifies the force (see (\ref{eq:force I}-\ref{eq:force V})). The large $T$ limit leads to an equivalence of ensembles between the microcanonical ensemble (where conditioning is on a fixed value of $Q_T$, defined in \eqref{eq:QT} and \eqref{eq:QT diff}) and the canonical ensemble (where the dynamics is weighted by $e^{\lambda\, Q_T}$). This is similar to the equivalence of thermodynamic ensembles in equilibrium when volume is large.

In the weak noise limit of the Langevin dynamics, one can introduce conditioned Large deviations functions which characterize fluctuations in the conditioned dynamics. Using a WKB solution we showed in \sref{sec:deformed ldf} that these conditioned large deviations functions can be expressed in terms of the solution of the Hamilton-Jacobi equations (\ref{eq:HJ1}-\ref{eq:HJ2}). The same result can also be derived (see \sref{sec:variational}) using a variational formulation, where the large deviations functions are related to the minimum of the Action that characterizes the path-space probability. Within this variational approach, one can calculate the optimal trajectory which describes how atypical fluctuations are generated and how they relax (\ref{eq:explicit z III},\,\ref{eq:sol z II}). A similar approach to our variational formulation was also used recently \cite{Nicolas2018} in the quasi-stationary regime of a Langevin dynamics in a periodic potential. 

One of the rather surprising aspects in the Langevin dynamics is that the noise becomes correlated over time due to the conditioning (see \eqref{eq:new noise correlation}). Moreover, fluctuations of the position at a time become correlated to the noise in the future.

The examples discussed in this paper are simple as they deal with a single degree of freedom. They are part of a theory which is rather general. In a forthcoming publication \cite{Sadhu2018} we shall apply the same ideas for a system with many degrees of freedom \cite{Bertini2014,Derrida2007,Hurtado2014}, \textit{e.g.} the symmetric exclusion process. The variational approach discussed here for the Langevin dynamics can be generalized for the large systems where the weak noise limit comes from the large volume. Several of the ideas used in this paper will be extended there.

We have seen in \eqref{eq:prob III} and \eqref{eq:pt Langevin III} that in the quasi-stationary regime the conditioned measure is a product of the left and right eigenvectors corresponding to the largest eigenvalue of the tilted matrix. Even in the non-stationary regime (see \eqref{eq:cond prob 2 app}) the conditioned measure is a product of a left vector and a right vector which evolve according to linear equations. This is very reminiscent of Quantum Mechanics. What could be learnt from this analogy is an interesting open issue.

\begin{acknowledgements}
We acknowledge the hospitality of ICTS-Bengaluru, India, where part of the work was completed during a workshop on Large deviations theory in August, 2017.
\end{acknowledgements}
\appendix

\section{Ensemble equivalence \label{app:equivalence}}
In this appendix we show that, for large $T$, the equivalence of ensembles holds for an arbitrary time $t$.

As the reasoning is very similar in the five regions of figure \ref{fig:fig1}, we will limit our discussion to the case of region II, \textit{i.e.} for $0\le t \ll T$. Let $P_t(C_T,C,Q \vert C_0)$ be the joint probability of configuration $C_T$ at time $T$, configuration $C$ at time $t$ and of the observable $Q_T$ to take value $Q$ given its initial configuration $C_0$ at time $0$. 

To establish the equivalence of ensembles in \eqref{eq:equivalence 2}, we need to show that the micro-canonical probability
\begin{equation}
\mathcal{P}_t(C \vert Q=qT)=\frac{\sum_{C_T}\sum_{C_0}P_t(C_T,C,Q=qT\vert C_0)R_{0}(C_0)}{\sum_{C'}\left[\sum_{C_T}\sum_{C_0}P_t(C_T,C',Q=qT\vert C_0)R_{0}(C_0)\right]}
\label{eq:P micro app}
\end{equation}
and the canonical probability 
\begin{equation}
P_t^{(\lambda)}(C)=\frac{\sum_{C_T}\sum_{C_0}\int dQ\, e^{\lambda Q}P_t(C_T,C,Q\vert C_0)R_{0}(C_0)}{\sum_{C'}\left[\sum_{C_T}\sum_{C_0}\int dQ\, e^{\lambda Q}P_t(C_T,C',Q\vert C_0)R_{0}(C_0)\right]}
\end{equation}
converge to the same distribution for large $T$ when $\lambda$ and $q$ are related by \eqref{eq:legendre}.

For this we write in terms of the probability \eqref{eq:joint P},
\begin{equation}
P_t(C_T,C,Q=qT\vert C_0)=\int dQ_t\;P_{T-t}(C_T,qT-Q_t \vert C)\;P_t(C,Q_t \vert C_0)
\end{equation}
and use the result \eqref{eq:PT large dev}. For large $T$, one has
\begin{equation*}
P_{T-t}(C_T,qT-Q_t \vert C)\simeq e^{-(T-t)\phi(q)-(tq-Q_t)\phi'(q)}\sqrt{\frac{\phi''(q)}{2\pi T}}R_{\phi'(q)}(C_T)L_{\phi'(q)}(C)
\end{equation*}
Substituting in \eqref{eq:P micro app} and simplifying the expression for large $T$ we get the micro-canonical probability
\begin{equation}
P_t(C \vert Q=qT)\simeq\frac{L_{\phi'(q)}(C)\sum_{C_0}G_t^{(\phi'(q))}(C\vert C_0)\;R_{0}(C_0)}{\sum_{C'} L_{\phi'(q)}(C')\sum_{C_0}G_t^{(\phi'(q))}(C'\vert C_0)\;R_{0}(C_0)}
\end{equation}
where $G_t^{(\lambda)}(C \vert C_0)$ is defined in \eqref{eq:G}.
On the other hand, using \eqref{eq:main G} for large $T$ we get the canonical probability
\begin{equation}
P_t^{(\lambda)}(C)\simeq\frac{L_{\lambda}(C)\sum_{C_0}G_t^{(\lambda)}(C\vert C_0)\;R_{0}(C_0)}{\sum_{C'} L_{\lambda}(C')\sum_{C_0}G_t^{(\lambda)}(C'\vert C_0)\;R_{0}(C_0)}
\label{eq:ptl}
\end{equation}
Clearly the two probabilities in the two ensembles coincide for $\lambda=\phi'(q)$.
Replacing $G_t^{(\lambda)}(C\vert C_0)$ by $M_{\lambda}^t(C,C_0)$ in \eqref{eq:ptl} leads to the conditioned measure \eqref{eq:prob II}. 

The same reasoning can be easily adapted in the other regions of \fref{fig:fig1}.

\section{Continuous time Markov process \label{app:cont time}}

\begin{figure}
\centering \includegraphics[width=0.6\textwidth]{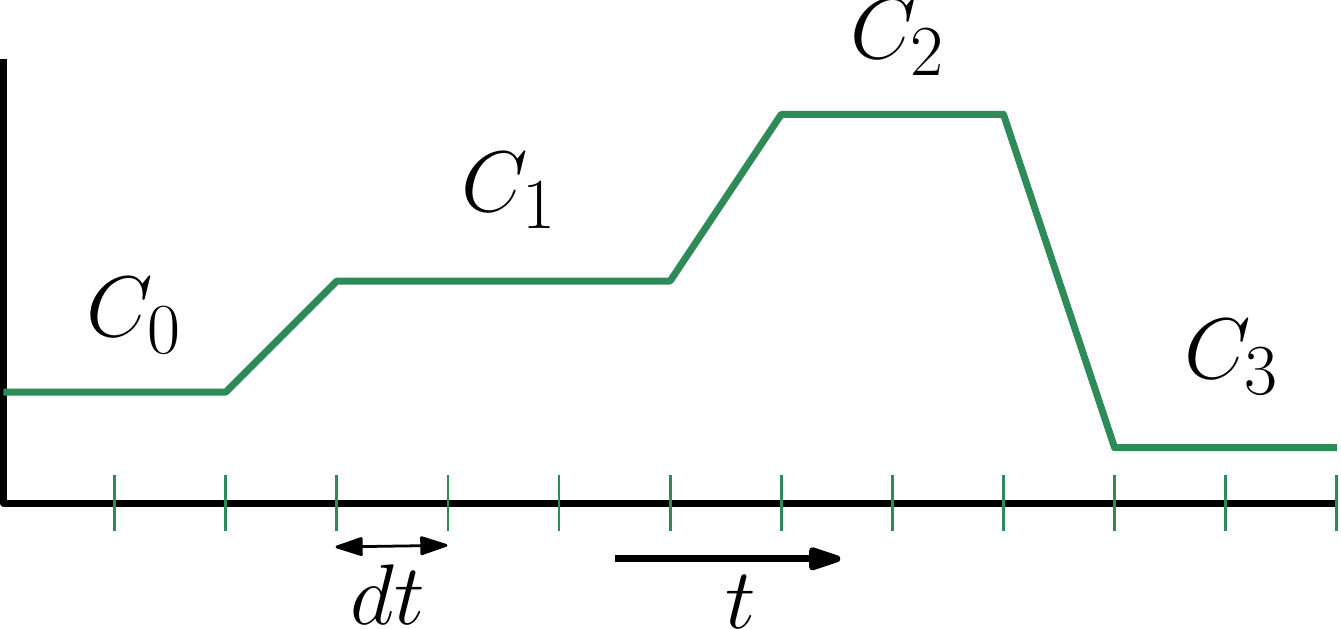}
\caption{A schematic of a time evolution in a Markov process with discrete time steps $dt$. The continuos time limit is obtained by taking $dt\rightarrow 0$ limit. \label{fig:conttimeevo}}
\end{figure}

In this Appendix, we describe a continuous time limit of the Markov process, illustrated in \fref{fig:conttimeevo}. In this, the empirical observable analogous to \eqref{eq:QT} is the $dt\rightarrow 0$ limit of
\begin{equation}
Q_T=dt\sum_{i=0}^{\frac{T}{dt}-1}f(C_i)+\sum_{n}g(C_n^+,C_n^-)
\label{eq:QT cont time app}
\end{equation}
where $t=i\, dt$, and $(C_n^-, C_n^+)$ are the configurations before and after the $n$th jump during the time interval $[0,T]$. 

From \eqref{eq:GT recursion} we get
\begin{equation*}
G_{T}^{(\lambda)}(C' \vert C_0)=\sum_{C}M_{\lambda}(C',C)\,G_{T-dt}^{(\lambda)}(C\vert C_0)
\end{equation*}
where
\begin{equation*}
M_\lambda(C',C)=\begin{cases} M_0(C',C)e^{\lambda [dt\, f(C)+g(C',C)]} & \textrm{for $C'\ne C$,}\cr M_0(C,C)e^{\lambda \, dt \, f(C)} & \textrm{for $C'= C$.}\end{cases}
\end{equation*}
Using the construction \eqref{eq:limit construction} for $M_0(C',C)$ we take the continuous time limit $dt\rightarrow 0$ and get
\begin{equation}
\frac{d}{dT} G_T^{(\lambda)}(C'\vert C_0)=\sum_{C}\mathcal{M}_{\lambda}(C',C)G_T^{(\lambda)}(C\vert C_0)
\label{eq:GT evo cont time}
\end{equation}
where $\mathcal{M}_{\lambda}$ is the tilted matrix for the continuous time process,
\begin{equation}
\mathcal{M}_{\lambda}(C',C)=\begin{cases} e^{\lambda g(C',C)}\mathcal{M}_0(C',C) & \textrm{for } C'\ne C,\\ \lambda f(C)-\sum_{C^{''}\ne C}\mathcal{M}_0(C^{''},C) & \textrm{for } C'= C. \end{cases}
\label{eq:tilted cont t}
\end{equation}
This shows that the generating function is the $(C,C_0)$th element of $e^{T\mathcal{M}_{\lambda}}$, \textit{i.e.}
\begin{equation}
G_T^{(\lambda)}(C\vert C_0)=e^{T\mathcal{M}_{\lambda}}(C,C_0)
\label{eq:GT Markov}
\end{equation}
Although \eqref{eq:GT evo cont time} resembles a Master equation, the tilted matrix $\mathcal{M}_{\lambda}$ is not a Markov matrix as $\sum_{C'} \mathcal{M}_{\lambda}(C',C)$ does not necessarily vanish.

For large $T$, one would get $G_{T}^{(\lambda)}(C \vert C_0)\simeq e^{T \mu(\lambda)}R_{\lambda}(C)L_{\lambda}(C_0)$ where the cumulant generating function $\mu(\lambda)$ is the largest eigenvalue of $\mathcal{M}_\lambda$ with $L_{\lambda}(C)$ and $R_\lambda(C)$ being the left and right eigenvectors, respectively. (Note the difference with the discrete time case \eqref{eq:main G} where $\mu(\lambda)$ is the \textit{logarithm} of the largest eigenvalue of the tilted matrix $M_\lambda$ in \eqref{eq:tilted discrete}.)

In a similar construction, one could get the continuous time limit of the conditioned measure \eqref{eq:prob I}-\eqref{eq:prob IV} and its time evolution \eqref{eq:tr prob 1}-\eqref{eq:tr prob 2}.
The analysis is straightforward and we present only the final result.

The time evolution of conditioned measure $P_t^{(\lambda)}(C)$ for a continuous time  Markov process is also a Markov process
\begin{equation}
\frac{d}{dt} P_t^{(\lambda)}(C')=\sum_C\mathcal{W}_t^{(\lambda)}(C',C)P_t^{(\lambda)}(C)
\label{eq:app Pt time evo cont}
\end{equation}
where $\mathcal{W}_t^{(\lambda)}(C',C)$ is the transition rate from $C$ to $C'$ at time $t$ in the canonical ensemble.
The conditioned measure and transition rate have different expressions in the five regions indicated in \fref{fig:fig1}. Their expression is given below where we use a matrix product notation, \textit{e.g.} $[L_\lambda \mathcal{M}_\lambda](C)\equiv \sum_{C'}L_\lambda(C') \mathcal{M}_\lambda(C',C)$.
\begin{enumerate}
\item Region I.
\begin{subequations}
\begin{align}
P_t^{(\lambda)}(C)=&\frac{[L_{\lambda}e^{-t\mathcal{M}_0}](C)R_{0}(C)}{\sum_{C'}L_{\lambda}(C')R_{0}(C')} 
\label{eq:app Pt time evo contn I}\\
\mathcal{W}_t^{(\lambda)} (C',C)=& {[L_\lambda e^{-t\mathcal{M}_0}](C') \over 
 [L_\lambda e^{-t\mathcal{M}_0}](C)  }\mathcal{M}_0(C',C)  -{\left[L_\lambda e^{-t\mathcal{M}_0}\mathcal{M}_0\right](C) \over 
 [L_\lambda e^{-t\mathcal{M}_0}](C)  }\;\delta_{C',C}
\end{align}
\end{subequations}
\item Region II.
\begin{subequations}
\begin{align}
P_t^{(\lambda)}(C)& =\frac{L_{\lambda}(C)[e^{t\mathcal{M}_\lambda}R_0](C)}{e^{t\mu(\lambda)}\sum_{C'}L_{\lambda}(C')R_{0}(C')}  \label{eq:app pt II}\\
\mathcal{W}_t^{(\lambda)}(C',C) &= { L_\lambda(C')    \over 
 L_\lambda(C)   }\mathcal{M}_\lambda(C',C)-\mu(\lambda)\delta_{C',C}   \label{eq:app w II}
\end{align}
\end{subequations}
\item Region III.
\begin{subequations}
\begin{align}
P_t^{(\lambda)}(C)& =L_{\lambda}(C)R_\lambda(C)\\
\mathcal{W}_t^{(\lambda)}(C',C) &= { L_\lambda(C')   \over 
 L_\lambda(C)   }\mathcal{M}_\lambda(C',C) -\mu(\lambda)\delta_{C',C} 
\end{align}
\end{subequations}
\item Region IV.
\begin{subequations}
\begin{align}
P_t^{(\lambda)}(C)=&\frac{[L_0\,  e^{(T-t)\mathcal{M}_\lambda}](C)R_{\lambda}(C)}{e^{(T-t)\mu(\lambda)}\sum_{C'}R_{\lambda}(C')}\\
\mathcal{W}_t^{(\lambda)}(C',C)= &{[L_0 \, e^{(T-t)\mathcal{M}_\lambda}](C')  \over 
 [L_0\,  e^{(T-t)\mathcal{M}_\lambda}](C)  }\mathcal{M}_\lambda(C',C)-  {  \left[ L_0\, e^{(T-t)\mathcal{M}_\lambda} \mathcal{M}_\lambda\right](C)\over 
 \left[ L_0\, e^{(T-t)\mathcal{M}_\lambda}\right](C) } \delta_{C',C}
\end{align}
\end{subequations}
where the left eigenvector $L_0$ for the original (unconditioned) evolution is a unit vector such that $[L_0\, \mathcal{M}_\lambda](C)\equiv \sum_{C'}\mathcal{M}_\lambda(C',C)$.
\item Region V.
\begin{subequations}
\begin{align}
P_t^{(\lambda)}(C)=&\frac{[e^{(t-T)\mathcal{M}_0}R_{\lambda}](C)}{\sum_{C'}R_{\lambda}(C')}
\label{eq:app Pt time evo cont V}\\
\mathcal{W}_t^{(\lambda)}(C',C)=&\mathcal{M}_0(C',C)
\end{align}
\end{subequations}
\end{enumerate}

One can verify the property $\sum_{C'}\mathcal{W}_t^{(\lambda)}(C',C)=0$ in all five regions. Moreover, setting $\lambda=0$, and $L_0(C)=1$, gives $\mathcal{W}_t^{(0)}(C',C)=\mathcal{M}_0(C',C)$, as one would expect.

\section{Conditioned Langevin dynamics \label{app:derivation FP}}
In this appendix, we show how the case of Langevin dynamics in \sref{sec:langevin} can be obtained as a continuous limit of the discrete time Markov process in \sref{sec:Markov}. One may alternatively derive the same results using the Kramers-Moyal expansion  \cite{VANKAMPEN2007193} of the continuous time Markov process in \aref{app:cont time}. 

\begin{figure}
\centering \includegraphics[width=0.8\textwidth]{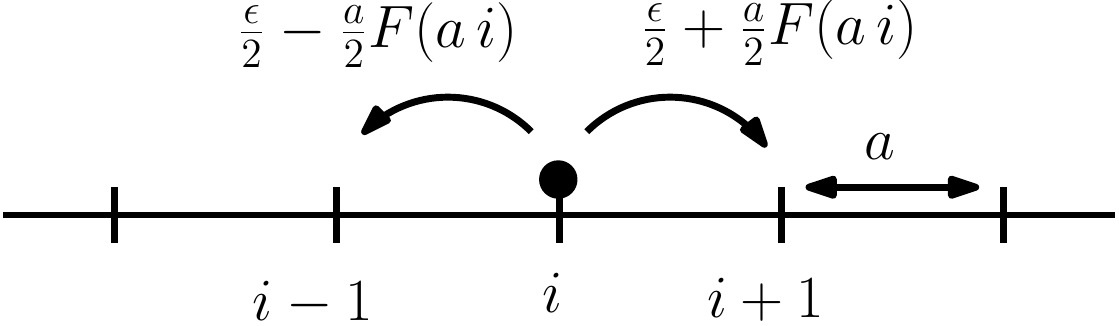}
\caption{A jump process on a one-dimensional chain where a particle jumps to its nearest neighbour site with rates indicated in the figure. \label{fig:jumpprocess}}
\end{figure}

In our approach, we consider a jump process on a one-dimensional lattice where a configuration $C$ is given by the site index $i$ as indicated in \fref{fig:jumpprocess}. Only nearest neighbor jumps are allowed with transition rate
\begin{equation}
M_0(i\pm 1,i)=\frac{\epsilon}{2}\pm \frac{a}{2}F(a\,i)
\label{eq:jump process}
\end{equation}
with $M_0(i,i)=1-\epsilon$, where $a$ is the unit lattice spacing, $\epsilon$ is a parameter, and $F(x)$ is an arbitrary function defined on the lattice. 

The probability  $P_{t,i}$  of the jump process to be in site $i$ at time $t$ satisfies
\begin{align}
P_{t+1,i}=M_0(i,i+1)P_{t,i+1}+M_0(i,i-1)P_{t,i-1}+M_0(i,i)P_{t,i}
\label{eq:discrete M eqn}
\end{align}
Taking the continuous limit $a\rightarrow 0$, one can easily see that, $P_{a^2 t}(a\, i)\equiv P_{t,i}$ follows the Fokker-Planck equation \eqref{eq:FP free}. This shows that the continuous limit of the jump process is indeed identical to the Langevin dynamics \eqref{eq:langvin free}.

One can now similarly derive results for the conditioned Langevin dynamics from the continuous limit of the jump process when it is conditioned to give a certain value of the observable $Q$ in \eqref{eq:Q general}. For this we define 
\begin{equation*}
f_t(i)=a^2 f(a\, i, a^2 t)\qquad  \textrm{and}\qquad g_t(i,j)=g(a\, i, a\, j, a^ 2 t)
\end{equation*}
Then, the continuous limit of \eqref{eq:Q general} corresponds to an observable $Q$ of the Langevin dynamics
\begin{equation}
Q=\int dt\, f(X_t,t)+\lim_{dt\rightarrow 0}\sum_{t}g(X_{t+dt},X_t,t)
\label{eq:Q diff general}
\end{equation}
Our choice $g_t[C,C]=0$ in \eqref{eq:Q general} leads to $g(x,x,t)=0$ at any time. This means, if we define $b_1(x,t)= \partial_x g(x,y,t)\vert_{y=x}$, $b_2(x,t)= \partial^2_xg(x,y,t)\vert_{y=x}$, $c_1(x,t)=\partial_yg(x,y,t)\vert_{y=x}$, and $c_2(x,t)=\partial^2_yg(x,y,t)\vert_{y=x}$ then
\begin{align}
b_1(x,t)+c_1(x,t)=0, \qquad \textrm{and} \qquad b_2(x,t)+ c_2(x,t)+2\partial_x\partial_y g(x,y,t)\vert_{y=x}=0
\label{eq:identity 0}
\end{align}
These identities also give 
\begin{equation}
-b_2(x,t)+c_2(x,t)+2b_1^{\prime}(x,t)=0
\label{eq:identity}
\end{equation}
which will be used in deriving some of the results below.

In the expression \eqref{eq:cond prob 2 app} for the conditioned measure if we define $H^{(\lambda)}_{a^2 t}(a\, i)\equiv Z_t^{(\lambda)}(i)$ and $\mathbb{H}^{(\lambda)}_{a^2 t}(a\, i)\equiv \mathbb{Z}_t^{(\lambda)}(i)$, then in the continuous limit $a\rightarrow 0$ we get the conditioned measure for the Langevin dynamics conditioned on \eqref{eq:Q diff general}:
\begin{equation}
P_t^{(\lambda)} (x)=\frac{ H^{(\lambda)}_t(x)\, \mathbb{H}^{(\lambda)}_t(x)}{\int dy\, H^{(\lambda)}_t(y) \, \mathbb{H}^{(\lambda)}_t(y)}
\label{eq:P general}
\end{equation}
The time evolution of $H^{(\lambda)}_t(x)$ and $\mathbb{H}^{(\lambda)}_t(x)$ are obtained from (\ref{eq:Z 1}-\ref{eq:Z 2}) for the jump process and taking $a\rightarrow 0$ limit. We get
\begin{subequations}
\begin{align}
\frac{d}{dt}H^{(\lambda)}_t & (x)=  \lambda f(x,t)H^{(\lambda)}_t(x)-\left(\frac{d}{dx}+\lambda\,c_1(x,t)\right)F(x)H^{(\lambda)}_t(x)\cr
 +\frac{\epsilon}{2}& \left(\frac{d^2}{dx^2}H^{(\lambda)}_t(x)+2\lambda\,c_1(x,t)\frac{d}{dx}H^{(\lambda)}_t(x)+\lambda\,c_2(x,t)H^{(\lambda)}_t(x)+\lambda^2 c_1(x,t)^2H^{(\lambda)}_t(x)\right)\label{eq:G eq app}\\
-\frac{d}{dt} \mathbb{H}^{(\lambda)}_t& (x)=  \lambda f(x,t)\mathbb{H}^{(\lambda)}_t(x)+F(x)\left(\frac{d}{dx}+\lambda\,b_1(x,t)\right)\mathbb{H}^{(\lambda)}_t(x)\cr
 + \frac{\epsilon}{2}&\left(\frac{d^2}{dx^2}\mathbb{H}^{(\lambda)}_t(x)+2\lambda\,b_1(x,t)\frac{d}{dx}\mathbb{H}^{(\lambda)}_t(x)+\lambda\,b_2(x,t)\mathbb{H}^{(\lambda)}_t(x)+\lambda^2 b_1(x,t)^2\mathbb{H}_t(x)\right) \label{eq:H eq app}
\end{align}
\end{subequations}

Similarly, using the identities (\ref{eq:identity 0}-\ref{eq:identity}), the continuous limit of (\ref{eq:Markov W},\,\ref{eq:W general}) gives the Fokker-Planck equation
\begin{subequations}
\begin{equation}
\frac{d}{dt}P_t^{(\lambda)}(x)=-\frac{d}{dx}\left[F_t^{(\lambda)}(x)P_t^{(\lambda)}(x) \right]+\frac{\epsilon}{2}\frac{d^2}{dx^2}P_t^{(\lambda)}(x)
\label{eq:FP general}
\end{equation}
where the modified force 
\begin{equation}
F_t^{(\lambda)}(x)=F(x)+\epsilon\left( \lambda\,b_1(x,t)+\frac{d}{dx}\log \mathbb{H}_t^{(\lambda)}(x)\right)
\label{eq:F general}
\end{equation}
\end{subequations}
This gives the time evolution of the Langevin dynamics when it is conditioned on the observable \eqref{eq:Q diff general}.

\begin{remarks}~\\
\begin{enumerate}
\item
In the derivation of \eqref{eq:FP general} one need use that the denominator in \eqref{eq:P general} is time independent which can be checked using (\ref{eq:G eq app},\,\ref{eq:H eq app}).
\item
The Fokker-Planck equation \eqref{eq:FP general} shows that the effect of conditioning a Langevin dynamics on an arbitrary time dependent observable \eqref{eq:Q diff general} is described by another Langevin dynamics with a modified force \eqref{eq:F general}, but the noise strength $\epsilon$ remains unchanged. This works even without a large parameter $T$ (see \cite{Orland2015} for an earlier example). 
\end{enumerate}
\end{remarks}

Our results in \sref{sec:langevin} belongs to a particular case, where the observable \eqref{eq:Q diff general} is defined in a large time interval $[0,T]$. This corresponds to
\begin{equation*}
f(x,t)=\begin{cases} f(x) \textrm{ for $t\in [0,T]$,}\cr  0 \textrm{ otherwise,}\end{cases} \quad g(x,y,t)=\begin{cases} (x-y)[\alpha h(x)+(1-\alpha)h(y)] \textrm{ for $t\in [0,T]$,}\cr  0 \textrm{ otherwise,}\end{cases}
\end{equation*}
and $T$ being large.
One can see that this corresponds to the observable \eqref{eq:QT diff}. In this case, (\ref{eq:G eq app},\,\ref{eq:H eq app}) gives
\begin{equation}
\frac{d}{dt}H^{(\lambda)}_t (x)=\mathcal{L}(t)\cdot H^{(\lambda)}_t (x),\qquad \frac{d}{dt}\mathbb{H}^{(\lambda)}_t (x)=-\mathcal{L}^\dagger(t)\cdot \mathbb{H}^{(\lambda)}_t (x)
\label{eq: H HH eqn}
\end{equation}
where $\mathcal{L}(t)=\mathcal{L}_\lambda$ for $t\in [0,T]$ and $\mathcal{L}(t)=\mathcal{L}_0$ outside this time window, with the 
operators defined in \eqref{eq:FP free} and \eqref{eq:tilted FP}; similar for the conjugate operator $\mathcal{L}^\dagger(t)$. This gives, for example, for $t\le 0$,  $H^{(\lambda)}_t (x)=r_0(x)$ (defined in \eqref{eq:eigen r l eqn}), whereas $\mathbb{H}^{(\lambda)}_t (x)\sim e^{-t\mathcal{L}_0^\dagger}\cdot e^{T\mathcal{L}_\lambda^\dagger}\cdot \ell_0(x)$, (upto a constant pre-factor) which in the large $T$ limit, gives $\mathbb{H}^{(\lambda)}_t (x)\sim e^{T\mu(\lambda)}\left[e^{-t\mathcal{L}_0^\dagger}\cdot \ell_\lambda\right](x)$. Substituting these results in \eqref{eq:P general} and \eqref{eq:F general} we get the expression for the conditioned measure \eqref{eq:pt Langevin I} and effective force \eqref{eq:force I}, respectively, in region I of \fref{fig:fig1}. For rest of the regions, the derivation is similar.

Lastly, from \eqref{eq:Z 1} one could see that for the observable \eqref{eq:QT}, the generating function $G_T^{(\lambda)}(C\vert C_0)$ in \eqref{eq:G} is identical to $Z_T^{(\lambda)}(C)$ if one sets $Z_0^{(\lambda)}(C)=\delta_{C,C_0}$. Then from the above calculation it is straightforward to show that in the continuous limit one would get \eqref{eq:GT recurrence cont}.

\section{Path integral formulation \label{app:Fok to Path}}

The path integral formulation of a Fokker-Planck equation is standard \cite{Kubo1973}.
The Fokker-Planck equation  \eqref{eq:FP free} can be written as
\begin{equation*}
\frac{dP_t(x)}{dt}=-\frac{d}{dx}\left[F(x)P_t(x)\right]+\frac{\epsilon }{2}\frac{d^2}{dx^2}P_t(x)\equiv -\mathcal{H}\left(x,-i \frac{d}{dx} \right)P_t(x)
\end{equation*}
such that $H(x,p)=F'(x)+iF(x)p+\frac{\epsilon}{2}p^2$. Considering a small increment $dt$ in time, we get
\begin{align*}
P_{t+dt}(x)\simeq & \int  dx'\left[1-dt\,\mathcal{H}\left(x,-i \frac{d}{dx} \right)\right]\delta(x-x') P_t(x') \cr \simeq & \int \frac{dp\, dx'}{2\pi}\left[1-dt\,\mathcal{H}\left(x,p \right)\right]e^{i\,p(x-x')} P_t(x')
\end{align*}
where we used the Fourier transform of the Dirac delta function $\delta(x-x')$. Iterating the evolution and taking $dt\rightarrow 0$ limit we get a path integral representation
\begin{equation*}
P_T(x)=\int_{z(0)=y}^{z(T)=x}\mathcal{D}[z,p]e^{\int_0^Tdt\, [ip\dot{z}-H(z,p)]}
\end{equation*}
with an initial condition $P_0(z)=\delta(z-y)$. The $H(z,p)$ is quadratic in $p$, and the corresponding path integral can be evaluated exactly, giving
\begin{equation*}
P_T(x)=\int_{z(0)=y}^{z(T)=x}\mathcal{D}[z]e^{-\frac{1}{2\epsilon}\int_0^Tdt\, (\dot{z}-F(z))^2-\int_0^Tdt\,F'(z)}
\end{equation*}
This is the path integral representation of the Fokker-Planck equation.

It is straightforward to generalize the above analysis  for the generating function \eqref{eq:GT recurrence cont}  and we get
 \begin{subequations}
 \begin{equation}
G_T^{(\lambda)}(x\vert y)=\int_{z(0)=y}^{z(T)=x}\mathcal{D}[z]e^{\mathbb{S}_T^{(\lambda)}[z(t)]}
\label{eq:G pathintegral}
\end{equation}
where the Action
\begin{equation}
\mathbb{S}_T^{(\lambda)}[z]=\int_0^T dt\left[\lambda  f(z)+\lambda \dot{z} h(z)-\frac{(\dot{z}-F(z))^2}{2\epsilon}-F'(z)-\epsilon \lambda \left(\alpha-\frac{1}{2}\right)h'(z)\right]
\label{eq:ST}
\end{equation}
\end{subequations}
Taking small $\epsilon$ limit, we get $\mathbb{S}_T^{(\frac{\kappa}{\epsilon})}[z]\simeq \frac{1}{\epsilon}S_T^{(\kappa)}[z]$ with the latter given in \eqref{eq:G path small} where we used $h(x)=0$.

\bibliographystyle{iopart-num}
\bibliography{reference,reference_old}

\providecommand{\newblock}{}
\begin{thebibliography}{10}
\expandafter\ifx\csname url\endcsname\relax
  \def\url#1{{\tt #1}}\fi
\expandafter\ifx\csname urlprefix\endcsname\relax\def\urlprefix{URL }\fi
\providecommand{\eprint}[2][]{\url{#2}}

\bibitem{Mey2014}
Mey A~S~J~S, Geissler P~L and Garrahan J~P 2014 {\em Phys. Rev. E\/} {\bf 89}
  032109

\bibitem{Delarue2017}
Delarue M, Koehl P and Orland H 2017 {\em J. Chem. Phys.\/} {\bf 147} 152703

\bibitem{Dykman1994}
Dykman M~I, Mori E, Ross J and Hunt P~M 1994 {\em J. Chem. Phys.\/} {\bf 100}
  5735

\bibitem{Laurie2015}
Lauri J and Bouchet F 2015 {\em N. J. Phys\/} {\bf 17} 015009

\bibitem{Garrahan2007}
Garrahan J~P, Jack R~L, Lecomte V, Pitard E, van Duijvendijk K and van Wijland
  F 2007 {\em Phys. Rev. Lett.\/} {\bf 98} 195702

\bibitem{Garrahan2009}
Garrahan J~P, Jack R~L, Lecomte V, Pitard E, van Duijvendijk K and van Wijland
  F 2009 {\em J. Phys. A\/} {\bf 42} 075007

\bibitem{Dorlas2001}
Dorlas T~C and Wedagedera J~R 2001 {\em Int. J. Mod. Phys. B\/} {\bf 15} 1

\bibitem{dembo2009}
Dembo A and Zeitouni O 2009 {\em Large Deviations Techniques and
  Applications\/} Stochastic Modelling and Applied Probability (Springer Berlin
  Heidelberg)

\bibitem{Varadhan1966}
Varadhan S~R~S 1966 {\em Communications on Pure and Applied Mathematics\/} {\bf
  19} 261

\bibitem{Varadhan2003}
Varadhan S~R~S 2003 {\em Communications on Pure and Applied Mathematics\/} {\bf
  56} 1222

\bibitem{Donsker19752}
Donsker M~D and Varadhan S~R~S 1975 {\em Communications on Pure and Applied
  Mathematics\/} {\bf 28} 1

\bibitem{Hollander2008}
den Hollander F 2008 {\em Large Deviations\/} Fields Institute monographs
  (American Mathematical Society)

\bibitem{TOUCHETTE2009}
Touchette H 2009 {\em Phys Rep\/} {\bf 478} 1

\bibitem{ellis1985}
Ellis R 1985 {\em Entropy, Large Deviations, and Statistical Mechanics\/} Die
  Grundlehren der mathematischen Wissenschaften in Einzeldarstellungen
  (Springer-Verlag)

\bibitem{Derrida2007}
Derrida B 2007 {\em J. Stat. Mech.\/}  P07023

\bibitem{Hurtado2014}
Hurtado P~I, Espigares C~P, del Pozo J~J and Garrido P~L 2014 {\em J. Stat.
  Phys.\/} {\bf 154} 214

\bibitem{Kurchan1998}
Kurchan J 1998 {\em J. Phys. A\/} {\bf 31} 3719

\bibitem{Gallavotti1995}
Gallavotti G and Cohen E~G~D 1995 {\em Phys. Rev. Lett.\/} {\bf 74} 2694

\bibitem{Lebowitz1999}
Lebowitz J~L and Spohn H 1999 {\em J. Stat. Phys.\/} {\bf 95} 333

\bibitem{Freidlin2012}
Freidlin M~I, Sz{\"u}cs J and Wentzell A~D 2012 {\em Random Perturbations of
  Dynamical Systems\/} Grundlehren der mathematischen Wissenschaften (Springer)

\bibitem{Graham1985}
Graham R and T\'el T 1985 {\em Phys. Rev. A\/} {\bf 31} 1109

\bibitem{Graham1973}
Graham R 1973 {\em Statistical Theory of Instabilities in Stationary
  Nonequilibrium Systems with Applications to Lasers and Nonlinear Optics\/}
  (Berlin, Heidelberg: Springer) p~1

\bibitem{Bertini2014}
Bertini L, De~Sole A, Gabrielli D, Jona-Lasinio G and Landim C 2015 {\em Rev.
  Mod. Phys.\/} {\bf 87} 593

\bibitem{Bertini2001}
Bertini L, De~Sole A, Gabrielli D, Jona-Lasinio G and Landim C 2001 {\em Phys.
  Rev. Lett.\/} {\bf 87} 040601

\bibitem{Varadhan1984}
Varadhan S~R~S 1984 {\em The Large Deviation Problem for Empirical
  Distributions of Markov Processes\/} (SIAM) p~33

\bibitem{Maes2008}
Maes C and Netocn\'{y} K 2008 {\em EPL (Europhysics Letters)\/} {\bf 82} 30003

\bibitem{Maes20082}
Maes C, Netocnn\'{y} K and Wynants B 2008 {\em Physica A\/} {\bf 387} 2675

\bibitem{Bodineau2004}
Bodineau T and Derrida B 2004 {\em Phys. Rev. Lett.\/} {\bf 92} 180601

\bibitem{Bertini2005Current}
Bertini L, De~Sole A, Gabrielli D, Jona-Lasinio G and Landim C 2005 {\em Phys.
  Rev. Lett.\/} {\bf 94} 030601

\bibitem{Hurtado2010}
Hurtado P~I and Garrido P~L 2010 {\em Phys. Rev. E\/} {\bf 81} 041102

\bibitem{Jack2010}
Jack R~L and Sollich P 2010 {\em Prog. Theo. Phys. Sup.\/} {\bf 184} 304

\bibitem{Jack2015}
Jack R~L and Sollich P 2015 {\em Euro. Phys. J. Special Topics\/} {\bf 224}
  2351

\bibitem{Chetrite2013}
Chetrite R and Touchette H 2013 {\em Phys. Rev. Lett.\/} {\bf 111} 120601

\bibitem{Chetrite2015}
Chetrite R and Touchette H 2015 {\em Ann. Henri Poincar{\'e}\/} {\bf 16} 2005

\bibitem{TOUCHETTE2017}
Touchette H 2018 {\em Physica A\/} {\bf 504} 5 lecture Notes of the 14th
  International Summer School on Fundamental Problems in Statistical Physics

\bibitem{Lecomte20072}
Lecomte V, Appert-Rolland C and van Wijland F 2007 {\em J. Stat. Phys.\/} {\bf
  127} 51

\bibitem{Hartmann2012}
Hartmann C and Sch\"{u}tte C 2012 {\em J. Stat. Mech.\/}  P11004

\bibitem{Bertini2015}
Bertini L, Faggionato A and Gabrielli D 2015 {\em Ann. Inst. H. Poincar\'{e}
  Prob. Stat.\/} {\bf 51} 867

\bibitem{Landau}
Landau L and Lifshitz E 1967 {\em Quantum Mechanics\/} (Moskow: MIR)

\bibitem{Evans2004}
Evans R~M~L 2004 {\em Phys. Rev. Lett.\/} {\bf 92} 150601

\bibitem{Stroock}
Strook D~W 2014 {\em {An Introduction to Markov Processes}\/} 2nd ed Graduate
  Texts in Mathematics (Springer)

\bibitem{Hirschberg2015}
Hirschberg O, Mukamel D and Sch{\"u}tz G~M 2015 {\em J. Stat. Mech.\/}  P11023

\bibitem{Schutz2016}
Sch{\"u}tz G~M 2016 {\em Duality Relations for the Periodic ASEP Conditioned on
  a Low Current\/} (Cham: Springer International Publishing) p 323

\bibitem{Popkov2011}
Popkov V and Sch{\"u}tz G~M 2011 {\em J. Stat. Phys.\/} {\bf 142} 627

\bibitem{Popkov2010}
Popkov V, Sch{\"u}tz G~M and Simon D 2010 {\em J. Stat. Mech.\/}  P10007

\bibitem{Fleming1992}
Fleming W~H 1992 Stochastic control and large deviations {\em Future Tendencies
  in Computer Science, Control and Applied Mathematics\/} ed Bensoussan A and
  Verjus J~P (Berlin, Heidelberg: Springer Berlin Heidelberg) p 291

\bibitem{Chetrite20152}
Chetrite R and Touchette H 2015 {\em J. Stat. Mech.\/}  P12001

\bibitem{Carollo2018}
Carollo F, Garrahan J~P, Lesanovsky I and P\'erez-Espigares C 2018 {\em Phys.
  Rev. A\/} {\bf 98} 010103

\bibitem{Roche2004}
Derrida B, Dou{\c{c}}ot B and Roche P~E 2004 {\em J. Stat. Phys.\/} {\bf 115}
  717

\bibitem{VANKAMPEN2007193}
Kampen N~v 2007 {\em Stochastic Processes in Physics and Chemistry (Third
  Edition)\/} third edition ed North-Holland Personal Library (Amsterdam:
  Elsevier)

\bibitem{Bodineau2005}
Bodineau T and Derrida B 2005 {\em Phys. Rev. E\/} {\bf 72} 066110

\bibitem{Baek2017}
Baek Y, Kafri Y and Lecomte V 2017 {\em Phys. Rev. Lett.\/} {\bf 118} 030604

\bibitem{Espigares2013}
Espigares C~P, Garrido P~L and Hurtado P~I 2013 {\em Phys. Rev. E\/} {\bf 87}
  032115

\bibitem{Sadhu2018}
Derrida B and Sadhu T 2018 {\em In preparation\/}

\bibitem{MckeanBook}
Mckean H~P 1969 {\em Stochastic Integrals\/} Probability and Mathematical
  Statistics: A Series of Monographs and Textbooks (Academic Press)

\bibitem{Sadhu2016}
Sadhu T and Derrida B 2016 {\em J. Stat. Mech.\/}  113202

\bibitem{Bertini2002}
Bertini L, Sole A~D, Gabrielli D and Landim C 2002 {\em J. Stat. Phys.\/} {\bf
  107} 635

\bibitem{Nicolas2018}
Nicolas T~E, Lecomte V and Bertin E 2018 {\em To appear\/}

\bibitem{Orland2015}
Majumdar S~N and Orland H 2015 {\em J. Stat. Mech.\/}  P06039

\bibitem{Kubo1973}
Kubo R, Matsuo K and Kitahara K 1973 {\em J. Stat. Phys.\/} {\bf 9} 51

\end{thebibliography}

\end{document}